\RequirePackage{fix-cm}
\documentclass[smallextended]{svjour3}       % onecolumn (second format)
\smartqed  % flush right qed marks, e.g. at end of proof
\usepackage{graphicx}
\usepackage{cite}
\usepackage{url}

\usepackage{fancyvrb}
\usepackage{xspace}
\usepackage{enumitem}

\usepackage[utf8]{inputenc}

\usepackage{amsmath}
\usepackage{amssymb}

\usepackage{mdframed}

\newcommand{\etal}{et al.\xspace}
\newcommand{\ie}{i.e.,\xspace}
\newcommand{\eg}{e.g.,\xspace}
\newcommand{\fig}[1]{Figure~\ref{#1}}
\newcommand{\tab}[1]{Table~\ref{#1}}
\newcommand{\sect}[1]{Section~\ref{#1}}

\newcommand{\cran}{\textsf{CRAN}\xspace}
\newcommand{\cpan}{\textsf{CPAN}\xspace}
\newcommand{\pypi}{\textsf{PyPI}\xspace}
\newcommand{\rubygems}{\textsf{RubyGems}\xspace}
\newcommand{\packagist}{\textsf{Packagist}\xspace}
\newcommand{\cargo}{\textsf{Cargo}\xspace}
\newcommand{\nuget}{\textsf{NuGet}\xspace}

\newcommand{\eclipse}{\textsf{Eclipse}\xspace}
\newcommand{\npm}{\textsf{npm}\xspace}
\newcommand{\github}{\textsf{GitHub}\xspace}
\newcommand{\git}{\textsf{Git}\xspace}

\newcommand{\javascript}{\textsf{JavaScript}\xspace}
\newcommand{\ruby}{\textsf{Ruby}\xspace}
\newcommand{\python}{\textsf{Python}\xspace}
\newcommand{\R}{\textsf{R}\xspace}
\newcommand{\perl}{\textsf{Perl}\xspace}
\newcommand{\php}{\textsf{PHP}\xspace}
\newcommand{\rust}{\textsf{Rust}\xspace}
\newcommand{\dotnet}{\textsf{.NET}\xspace}

\newcommand{\figsize}{0.9\textwidth}

\begin{document}

\title{An Empirical Comparison of Dependency Network Evolution in Seven Software Packaging Ecosystems} 


\titlerunning{A Comparative Study of Package Dependency Networks}        % if too long for running head

\author{Alexandre Decan         \and
        Tom Mens \and
        Philippe Grosjean
}


\institute{A. Decan \and T. Mens \and Ph. Grosjean \at
              COMPLEXYS Research Institute, University of Mons, Belgium \\
%              Tel.: +32 65 37 3453\\
              \email{\{ firstname . lastname \}@umons.ac.be}           %  \\
  %         S. Author \at
    %          second address
}


\maketitle

\begin{abstract}
Nearly every popular programming language comes with one or more package managers.
The software packages distributed by such package managers form large software ecosystems. These packaging ecosystems contain a large number of package releases that are updated regularly and that have many dependencies to other package releases.
While packaging ecosystems are extremely useful for their respective communities of developers,
they face challenges related to their scale, complexity, and rate of evolution.
Typical problems are backward incompatible package updates, and the risk of (transitively) depending on packages that have become obsolete or inactive.

This manuscript uses the \textsf{libraries.io} dataset to carry out a quantitative empirical analysis of the similarities and differences between the evolution of package dependency networks for seven packaging ecosystems of varying sizes and ages:  
\cargo for \rust, \cpan for \perl, \cran for \R, \npm for \javascript, \nuget for the \dotnet platform, \packagist for \php, and \rubygems for \ruby.

We propose novel metrics to capture the growth, changeability, resuability and fragility of these dependency networks, and use these metrics to analyse and compare their evolution.
We observe that the dependency networks tend to grow over time, both in size and in number of package updates,
while a minority of packages are responsible for most of the package updates.
The majority of packages depend on other packages, but only a small proportion of packages accounts for most of the reverse dependencies.
We observe a high proportion of ``fragile'' packages due to a high and increasing number of transitive dependencies.
These findings are instrumental for assessing the quality of a package dependency network,
and improving it through dependency management tools and imposed policies.
\end{abstract}

%%%%%%%%%%%%%%%%%%%%%%%%%%%%%%

\section{Introduction}\label{sec:intro}

Traditionally, software engineering research has focused on understanding and improving the development and evolution of individual software systems. The widespread use of online collaborative development solutions surrounding distributed version control tools (such as \git and \github) has lead to an increased popularity of so-called \emph{software ecosystems}, large collections of interdependent software components that are maintained by large and geographically distributed communities of collaborating contributors. Typical examples of open source software ecosystems are distributions for Linux operating systems and \emph{packaging ecosystems} for specific programming languages. 

Software ecosystems tend to be very large, containing from tens to hundreds of thousands of packages, with even an order of magnitude more dependencies between them.
Complicated and changing dependencies are a burden for many developers and are often referred to as the``dependency hell'' \cite{Artho2012, Bogart2016}. 
If not properly maintained, the presence of such dependencies may become detrimental to the ecosystem quality. Indeed, developers are reluctant to upgrade their dependencies \cite{Bavota2015}, while outdated dependencies have been shown to be more vulnerable to security issues \cite{Cox2015}. 
Researchers have therefore been actively studying the evolution dynamics of packaging dependency networks in order to support the many problems induced by their macro-level evolution \cite{Decan2016SANER,Gonzalez-Barahona2009}.
A famous example of such a problem was the \textsf{left-pad} incident for the \textsf{npm} package manager.
Despite its small size (just a few lines of source code), the sudden and unexpected removal of the \textsf{left-pad} package
caused thousands of direct and indirect dependent projects to break, including very popular ones such as \textsf{Atom} and \textsf{Babel} \cite{NPM2016,Haney2016}.

Comparative studies between package dependency networks of different ecosystems are urgently needed, to understand their similarities and differences and how these evolve over time. Such studies may help to improve soffware analysis tools by taking into account specific ecosystem characteristics to better manage and control the intrinsic fragility and complexity of evolving software ecosystems.

The current paper builds further upon our previous work. In \cite{Decan2016WEA} we empirically compared the package dependency network of three popular packaging ecosystems: \cran for \R, \pypi for \python and \npm for \javascript. These analyses focused on the structural complexity of these dependency networks. We found important differences between the considered ecosystems, which can partly be explained by the functionality offered by the standard library of the ecosystem's underlying programming language, as well as by other ecosystem specificities. This implies that the findings for one ecosystem cannot necessarily be generalised to another. We therefore suggested to extend the analyses by considering more ecosystems, and by taking into account the evolution of the dependency networks. 
In \cite{Decan2017SANER}, we started to carry out an historical analysis of package dependency network evolution for \cran, \npm, and \rubygems. We studied to which extent packages rely on other packages, as well as to which extent packages updates are problematic in the presence of (transitive) package dependencies. We observed that, because of the presence of many transitive dependencies, a package failure may potentially affect many other packages. 
While each ecosystem provides specific and different ways to reduce the impact of problematic package updates, none of these solutions are perfect and package maintainers remain faced with occasional update problems.

The current paper extends the results of \cite{Decan2017SANER} by considering seven different packaging ecosystems for as many different programming languages: \cargo for \rust, \cpan for \perl, \cran for \R, \npm for \javascript, \nuget for the \dotnet development platform, \packagist for \php, and \rubygems for \ruby. As far as we know, this is the first work to compare that many different ecosystems, and to use the recent l\textsf{ibraries.io} dataset for that purpose.
Another novelty is that we introduce three new metrics to facilitate ecosystem comparison despite the diversity of the considered ecosystems is terms of age and size: the \textit{Changeability Index} captures an ecosystem's propensity to change over time; the \textit{Reusability Index} captures the ecosystem's amplitude and extent of reuse; and the \textit{P-Impact Index} assesses the fragility of an ecosystem.

The remainder of this article is structured as follows.
\sect{sec:related} discusses related work.
\sect{sec:methodo} presents the used terminology, motivates the selected packaging ecosystems and explains the data extraction process.
Sections~\ref{sec:rq-growth} to \ref{sec:rq-complexity} each address a specific research question.

\sect{sec:rq-growth} studies our first research question: \emph{``How do package dependency networks grow over time?''}
We observe a continuing growth of the number of packages and their dependency relationships.
Given that we observed in \cite{Decan2017SANER} that package dependencies may be problematic in case of package updates, \sect{sec:rq-change} studies a second research question: \emph{``How frequently are packages updated?''}
Because package dependencies lead to an increased fragility of the ecosystem, \sect{sec:rq-reuse} studies a third research question \emph{``To which extent do packages depend on other packages?''}
\sect{sec:rq-complexity} studies the fourth research question: \emph{``How prevalent are transitive dependencies?''} Indeed, due to the prevalence of transitive dependencies in the package dependency network, package failures may propagate through the network and may impact large parts of the ecosystem.

\sect{sec:threats} presents the threats to validity of our study.
\sect{sec:discussion} puts our research findings into perspective, by discussing how an ecosystem's policy influences the observed results, what are the limitations of existing techniques to deal with package dependencies and package updates, and how our results could form the basis of ecosystem-level health analysis tools.
\sect{sec:futurework} outlines future work, by providing initial evidence for laws of software ecosystem evolution, and suggesting to explore software ecosystem evolution from a complex network or socio-technical network point of view.
Finally, \sect{sec:conclusion} concludes.

%%%%%%%%%%%%%%%%%%%%%%%%%%%%%%

\section{Related Work}\label{sec:related}

The research domain of software ecosystems is huge. We refer the reader to some recent key references for further reading \cite{ManikasHansen2012,Jansen2013softwareecosystems,SerebrenikMens2015}.
Given that the current article specifically focuses on \emph{packaging} ecosystems, and more in particular on technical dependencies in package \emph{dependency networks}, this section reports mainly on the related work in those areas. Although very interesting in their own right, social dependency networks are out of scope for the current work, and work related to such networks will therefore not be discussed here.

Many researchers have studied (and compared) technical dependency networks at the level of components contained within individual software projects (\eg studying the modularity of the dependency network between classes in a Java project \cite{Dietrich2008SoftVis,Zanetti2012ARCS}). A detailed account of such works is outside the scope of the current article, since our focus is at the ecosystem level, i.e., we consider dependencies across different projects (as opposed to within individual projects).

Many researchers have studied package dependencies issues in a variety of programming language packaging ecosystems. Most studies, however, were limited to a single ecosystem.
Wittern \etal~\cite{Wittern2016} studied the evolution of a subset of \javascript packages in \npm, analysing characteristics such as their dependencies, update frequency, popularity, version numbering and so on.
Abdalkareem~\etal~\cite{Abdalkareem2017ESECFSE} also carried out an empirical case study of \npm, focusing on what they refer to as ``trivial'' packages, and the risk of depending on such packages. The results were inconclusive, in the sence that depending on trivial packages can be useful and unrisky, provided that they are well implemented and tested.
The \cran packaging ecosystem has been previously studied~\cite{Hornik2012, GermanAdamsHassan2013,DecanMCG2015ecsa}, and dependencies have been shown to be an important cause of errors in \R packages both on \cran and \github~\cite{Decan2016SANER}.
Blincoe \etal \cite{Blincoe2015} looked at \ruby as part of a larger \github study on the emergence of software ecosystems, and observed that most ecosystems are centered around one project and are interconnected with other ecosystems. 
Bavota \etal~\cite{Bavota2015} studied the evolution of dependencies in the \textsf{Apache} ecosystem and highlighted that dependencies have an exponential growth and must be taken care of by developers. 
Considering that changes of a package might break its dependent packages, Bavota \etal found that developers were reluctant to upgrade the packages they depend on.
Robbes \etal~\cite{Robbes2012-smalltalkecos} studied the ripple effect of API method deprecation in the \textsf{Smalltalk} ecosystem and revealed that API changes can have a large impact on the system and remain undetected for a long time after the initial change.

Santana~\etal~\cite{Santana2013IWSECO} focused on the visualisation aspects of software ecosystem analysis, and proposed a social visualisation of the  interaction between contributors of the community on the one hand, and a technical visualisation of the ecosystem's project dependencies on the other hand. They did not focus, however, on how to compute or visualise metrics about the ecosystem.

The dependence on packages with security vulnerabilities has been studied in industrial software projects \cite{Cadariu2015}. Cox \etal revealed that systems using outdated dependencies are four times more likely to have security issues as opposed to systems that are up-to-date \cite{Cox2015}.

Very little research results are available that actually compare dependency and maintainability issues across different packaging ecosystems.
Bogart \etal \cite{Bogart2016} compared three ecosystems (\npm, \cran and \eclipse) in order to understand the impact of community values, tools and policies on breaking changes. They carried out a qualitative analysis by relying on interviews with developers of the studied ecosystems. Specifically related to package dependencies, they identified two main types of mitigation strategies adopted by package developers to reduce their exposure to changes in other packages: limiting the number of dependencies; and selecting only dependencies to packages that they trust.
Bogart's work complements the current paper, which is based on a quantitative empirical comparison of the dependency networks of packaging ecosystems. 

Inspired by our own previous work \cite{Decan2016WEA, Decan2017SANER}, Kikas~\etal~\cite{Kikas2017MSR} carried out an empirical comparison of the dependency networks of three ecosystems (\npm, \rubygems and \rust), confirming our own findings related to the ecosystems' fragility and vulnerability to transitive dependencies.

%%%%%%%%%%%%%%%%%%%%%%%%%%%%%%%%%%%%%%%%%%
\section{Methodology}\label{sec:methodo}

\subsection{Preliminaries}

All empirical analysis presented in the current article is supported by a replication package, available on GitHub as Python notebooks\footnote{See \url{https://github.com/AlexandreDecan/ecos-emse}}.

\tab{tab:terminology} informally defines all terms used in this article. The parts of the term indicated between parentheses in the first column of the table will be implicitly assumed if they are clear from the context. 

\begin{table}[!htbp]
\caption{Informal definition of terms used in this article.}
\label{tab:terminology}\centering
\begin{tabular}{p{1.85cm}|p{9.15cm}}
  \bf Term & \bf Informal definition \\ \hline \hline
  (Packaging) \bf Ecosystem & The collection and history of all tools, software artefacts and community members surrounding a particular \emph{package manager}.\\  \hline
  \bf Package Manager & A coherent collection of software tools that automates the process of installing, configuring, upgrading or removing software \emph{packages} on a computer's operating system in a consistent manner.  \\  \hline
  \bf Package & A computer program providing specific functionalities. A package usually exists in many versions which are called \emph{release}s. By abuse of language, a \emph{package} at time $t$ denotes its latest available \emph{release} at time $t$. \\  \hline
  (Package)\quad \bf Release & A specific version of a \emph{package} that can be accessed and installed through the \emph{package manager}. It usually comes in the form of an archive file containing what is needed to build, configure and deploy the package version, and includes a \emph{manifest} containing important metadata such as its owner, name, description, timestamp, and a list of direct \emph{dependencies} to other \emph{packages} that are required for its proper functioning.\\  \hline
  (Package)\quad \bf Update & A new \emph{release} of a package, provided by the \emph{package manager},  that succeeds (\ie corresponds to a higher version number or timestamp) a previous release of the same package. \\  \hline
  (Package)\quad\quad {\bf Dependency Network}\quad (at time $t$) & A graph structure in which the nodes represent all the \emph{packages} made available by the \emph{package manager} at time $t$, and the directed edges represent \emph{direct dependencies} between the latest available \emph{releases} at time $t$.\\  \hline
 \bf Dependency & An explicitly documented reference (in the manifest of a \emph{release}) to another \emph{package} that is required for its proper functioning. A dependency can specify constraints to restrict the supported \emph{releases} of the target package. Dependencies that are explicitly documented in the release manifest (\ie edges in the \emph{dependency network}) are called \textbf{direct dependencies}. Those that are part of the transitive closure of the \emph{dependency network} are called \textbf{transitive dependencies}. Transitive dependencies that are not direct are called \textbf{indirect dependencies}. \\  \hline
 \bf Reverse \quad Dependency & Reverse dependencies are obtained by following the edges of the \emph{dependency network} in the opposite direction. As for normal dependencies, they can be \textbf{direct}, \textbf{transitive} or \textbf{indirect}. \\  \hline
  {\bf Required} \quad package & A \emph{package} that is the target of at least one \emph{dependency} from another \emph{package}. In a similar vein, we define {\bf transitively required}.\\  \hline
  {\bf Dependent} \quad (package) & A \emph{package} that is the target of at least one \emph{reverse dependency} from another \emph{package}. In a similar vein, we define {\bf transitively dependent}. \\  \hline
  {\bf Connected} \quad package & A package that is either a \emph{required} or a \emph{dependent} package. \\  \hline
  {\bf Top-level} \quad package & A \emph{dependent package} that is not a \emph{required} package.
  \end{tabular}
\end{table}

%%%%%%%%%%%%%%%%%%%%%%%%%%%%%%
\subsection{Statistical analysis techniques}

One of the statistical techniques that will be used in this article is \emph{survival analysis} (a.k.a. event history analysis)~\cite{Aalen2008}.
It is a technique that models ``time to event'' data with the aim to estimate the survival rate of a given population, \ie the expected time duration until a specific ``event'' happens (such as death of a biological organism, failure of a mechanical component, recovery of a disease). A common non-parametric statistic used to estimate survival functions is the Kaplan-Meier estimator \cite{KaplanMeier2012}.

Survival analysis models take into account the fact that some observed subjects may be ``censored'', either because they leave the study during the observation period, or because the event of interest was not observed on them during the observation period.
In empirical software engineering, survival analysis has been used to estimate the survival of open source projects over time \cite{Samoladas2010}, to analyse the use and removal of functions in PHP code \cite{Kyriakakis2014}, to analyse dead Java code \cite{Scanniello2011}, to analyse the survival of database access libraries in Java code \cite{GoeminnetEtAl2015-ICSME,Decan2017-CEUR}, and to analyse survival of developers in open source projects \cite{Lin2017}. Inspired by this research, in this paper we will use the technique to analyse the survival of package releases in packaging ecosystems.

\bigskip

As several research questions require to measure statistical dispersion, we borrowed ideas from econometrics and used the Lorenz curve~\cite{doi:10.1080/15225437.1905.10503443} and the related Gini index~\cite{Gini1912}. 
Those two techniques are usually applied to assess the inequality of the wealth distribution among people, regions, countries, and so on. 
The Lorenz curve is typically used to compare graphically the cumulative proportion of income versus the cumulative proportion of individuals, illustrating the inequality of a wealth distribution. 
The Gini coefficient (or Gini index) is a widely used social and economic indicator to cope with unevenly distributed data. 
Its value is comprised between 0 and $1 - \frac{1}{n}$, where $n$ is the size of the considered population.
A value of 0 expresses perfect equality and a value of $1 - \frac{1}{n}$ expresses maximal inequality among individuals, where one individual possesses all of the wealth of the given population.

Gini index has been previously used in empirical software engineering. 
Considering software metrics data as wealth distributions, Vasa~\etal~\cite{Vasa2009Gini} showed that many software metrics not only display high Gini values, but that these values are remarkably consistent over time. 
Giger~\etal~\cite{Giger2011} used the index to investigate how changes made to source code are distributed in the \textsf{Eclipse} project.
Goeminne~\etal~\cite{Goeminne2011-SQM} measured the inequality of different kinds of activity in open source software projects using different econometrics, including Gini, and found empirical evidence of highly skewed distributions in the activity of developers involved in open source software projects.

%%%%%%%%%%%%%%%%%%%%%%%%%%%%%%
\subsection{Selected Packaging Ecosystems}

This article focuses on programming language ecosystems, and more specifically packaging ecosystems revolving around package managers for specific programming languages. Such ecosystems tend to have a very active community of contributors, making their dependency networks very large, and causing difficulties in managing and analysing the evolution of these networks.
Given that these ecosystems serve a similar goal, namely to serve the developer community surrounding a particular programming language, it makes sense to empirically compare them.

\begin{table}[!htbp]
\caption{Characteristics of the selected packaging ecosystems on 1 April 2017: creation year, language, number of packages, number of releases, number of dependencies across all releases, and release date of the oldest package.}
\label{tab:characteristics}\centering
\begin{tabular}{l||c|c|r|r|r|c}
  \bf Manager & \bf Creation & \bf Lang. & \bf Pkg. & \bf Rel. & \bf Deps. & \bf Oldest pkg.\\ \hline
  \bf \cargo & 2014 & \rust & 9k & 48k & 150k & 2014-11 \\
  \bf \cpan & 1995 & \perl & 34k & 259k & 1,078k & 1995-08 \\
  \bf \cran & 1997 & \R & 12k & 67k & 164k & 1997-10 \\
  \bf \npm & 2010 & \javascript & 462k & 3,038k & 13,695k & 2010-11 \\
  \bf \nuget & 2010 & \dotnet & 84k & 936k & 1,665k & 2011-01 \\
  \bf \packagist & 2012 & \php & 97k & 669k & 1,863k & 1994-08 \\
  \bf \rubygems & 2004 & \ruby & 132k & 795k & 1,894k & 1999-12 %\\
\end{tabular}
\end{table}

The seven ecosystems we selected form a representative collection of package managers, covering different programming languages, dependency network sizes and ages, as summarized in Table~\ref{tab:characteristics}.
On 1 April 2017, these ecosystems hosted together 5,812k releases for more than 830k packages. Among those package releases, we identified 20,509k dependency relationships.
A brief description of each considered package manager is presented below:

\begin{itemize}
\item \cargo is the official package manager for \rust, a compiled programming language released in 2012 by Mozilla. Since 2014, its official package registry is \url{crates.io}, usually referred to as \cargo. It is the youngest and smallest of the selected ecosystems. 

\item \cpan (\url{cpan.org}) stands for Comprehensive Perl Archive Network and is the oldest considered ecosystem. It was introduced in 1995 as a large collection of \perl software, an interpreted programming language developed in 1987. 

\item \cran (\url{cran.r-project.org}), the Comprehensive R Archive Network, is the second oldest ecosystem we consider. It constitutes the official repository of the statistical computing environment \R. It has the particularity of following a ``rolling release'' policy, meaning that only the latest release of a package can be automatically installed from \cran. As a consequence, packages must always be compatible with the latest release of each of their dependencies, as well as with the latest version of the \R language. 

\item \npm (\url{npmjs.com}), started in 2010, is the official package registry for the \javascript runtime environment \textsf{Node.js}. It is the largest considered ecosystem with nearly half a million packages.

\item \nuget (\url{nuget.org}), formerly known as \textsf{NuPack}, is the official package manager developed by Microsoft for the \dotnet development platform.  
By extension, \nuget also designates \textsf{NuGet Gallery}, the central package repository for \nuget. 

\item \packagist (\url{packagist.org}) is the default package repository for \textsf{Composer}, the de-facto standard package manager for the interpreted, web-oriented programming language \php. Although \packagist was started in 2012, it also hosts packages that were developed prior to its release, in the early days of \php (1994).

\item \rubygems (\url{rubygems.org}) is the largest collection of packages for \ruby, an interpreted object-oriented programming language. \rubygems was started on \textit{Pi Day} 2004 and, like \packagist, also hosts packages that were developed prior to its release.
\end{itemize}

Unless explicitly mentioned, all conducted statistical analyses considered the whole lifetime of each ecosystem up to 1 January 2017. 
In the accompanying figures we decided to display only the period starting from 1 January 2012 to 1 January 2017 for the sake of clarity. 

%%%%%%%%%%%%%%%%%%%%%%%%%%%%%%
\subsection{Data Extraction Process}

For our empirical study, we relied on information about package releases and dependencies collected by the open source discovery service \textsf{libraries.io}\footnote{\url{https://libraries.io} ; \url{https://zenodo.org/record/808273}}. 
The extracted data falls under the CC-BY-SA 4.0 licence.\footnote{Creative Commons Attribution-ShareAlike 4.0 International,\\ see \url{https://creativecommons.org/licenses/by-sa/4.0/}.}
\textsf{libraries.io} extracted all the metadata from the manifest of each package, based on the list of packages provided by the official registry of the packaging manager.

At the time of carrying out our experiment, \textsf{libraries.io} provided package release data for 33 popular package managers in total.
We excluded those that we considered too small (less than 5,000 packages). From the remaining 17 packaging ecosystems, we selected seven for which it was possible to obtain all the necessary metadata from the package manifests statically: \cargo, \cpan, \cran, \npm, \nuget, \packagist, and \rubygems.

We excluded the other packaging ecosystems from our study for a variety of reasons:
because they were too domain-specific, targeting a specific software framework (\eg \textsf{Meteor}) or software component (\eg \textsf{WordPress} and \textsf{Atom}); because they host a subset of packages available through another already considered ecosystem (\eg \textsf{Bower} manages a subset of \npm); or because the developers of \textsf{libraries.io} informed us that important dependency information was incomplete or missing (\eg \textsf{GO, PyPi, Maven, CocoaPods, Clojars} or \textsf{Hackage}).
The latter case applied to two very popular and important packaging ecosystems, namely \textsf{Maven} for Java and \textsf{PyPi} for Python. For these package managers,  (the list of) package dependencies can be dynamically defined and may depend on the environment that interprets the manifest at installation time, and hence are not available statically. An interesting topic of future work would therefore constitute the automated analysis and extraction of such dynamic dependencies.

To ascertain the correctness of the data provided by \textsf{libraries.io}, we manually cross-checked its retrieved dependency metadata with our own (less recent) datasets for \cran, \npm and \rubygems that we had used in our previous work \cite{Decan2017SANER}. The metadata matched for the considered period, convincing us of its correctness. 

For \cran, we completed the metadata extracted from \textsf{libraries.io} with data about archived package releases (\ie releases that used to be distributed on \cran but are no longer available through the package manager). To achieve this, we relied on \textsf{extractoR}, a publicly available\footnote{\url{https://github.com/ecos-umons/extractoR}} \R package that was developed specifically for the purpose of mining and analysing \cran packages \cite{ClaesMG2014}. 
With the help of \textsf{extractoR}, we retrieved the metadata of 1,078 additional packages and 5,182 additional package releases.

For \npm, we observed that the \textsf{left-pad} incident~\cite{NPM2016,Haney2016} seems to have lead some developers to design packages whose sole purpose is to depend on as many other packages as possible. For instance, the \textsf{npm-gen-all} package is defined as a package that ``\emph{will create a multitude of \npm projects that will depend on every \npm package published}''. We identified around 250 of such packages, and explicitly ignored them for our analyses since they only introduce noise and do not serve any useful purpose. 

For each package release of each considered packaging ecosystem, we considered the list of packages on which it depends. 
We restricted the dependencies to those required to install and execute the package. Dependencies that are only required to develop or test a package were excluded from our analyses because not all ecosystems make use of them. Even for ecosystems that support them, not every package declares a complete and reliable list of development or test dependencies.
Depending on the ecosystem, this means that we restricted ourselves to dependencies of type ``runtime'', ``imports'', ``depends'' and ``normal'', while omitting dependencies of type ``development'', ``optional'', ``enhances'', ``suggests'', ``build'', ``configure'', ``test'', ``develop'' or ``dev''.
We also excluded dependencies that target packages that were not available through the package manager (\eg packages that are hosted directly on the web or on \git repositories). This represents less than 2.5\% of all dependencies in \cargo, \cran, \npm, \nuget and \rubygems, around 9.35\% of the dependencies in \packagist and 30.11\% of those in \cpan. A possible explanation for this higher proportion of unavailable dependencies in \cpan relates to the presence of very old packages that are not maintained anymore and still depend on packages that are no longer available on \cpan.

%%%%%%%%%%%%%%%%%%%%%%%%%%%%%%%%%%%%%%%%%%
\section{$RQ_1$: How do package dependency networks grow over time?}\label{sec:rq-growth}

As a first research question, we study how fast each packaging ecosystem and package dependency network is growing over time. Being aware of this speed of growth is important, since it may become increasingly difficult to manage the ecosystem without putting in place proper policies, processes, quality standards and tool support capable of managing this growth \cite{Hornik2012}.
Our hypothesis is that the number of new packages must continuously increase in order to offer new functionality to the ecosystem users. On the other hand, the increase should not be too fast, since a higher number of dependencies between packages makes the ecosystem more interconnected and therefore more complex.

\begin{figure}[!htbp]
   \centering
   \includegraphics[width=\figsize]{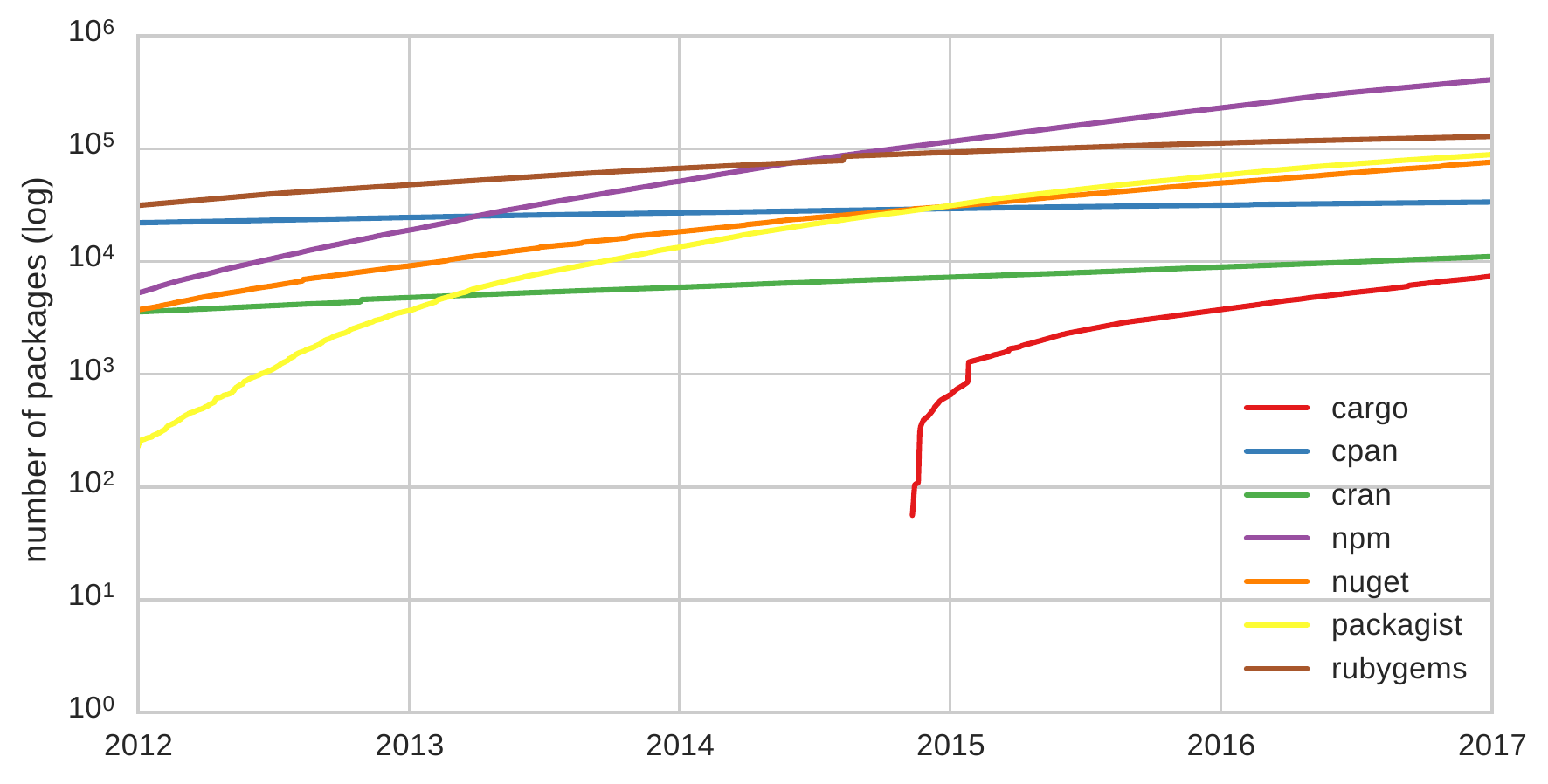}
   \caption{
     Evolution of the number of packages.
   }
   \label{fig:act_number_of_packages}
  \includegraphics[width=\figsize]{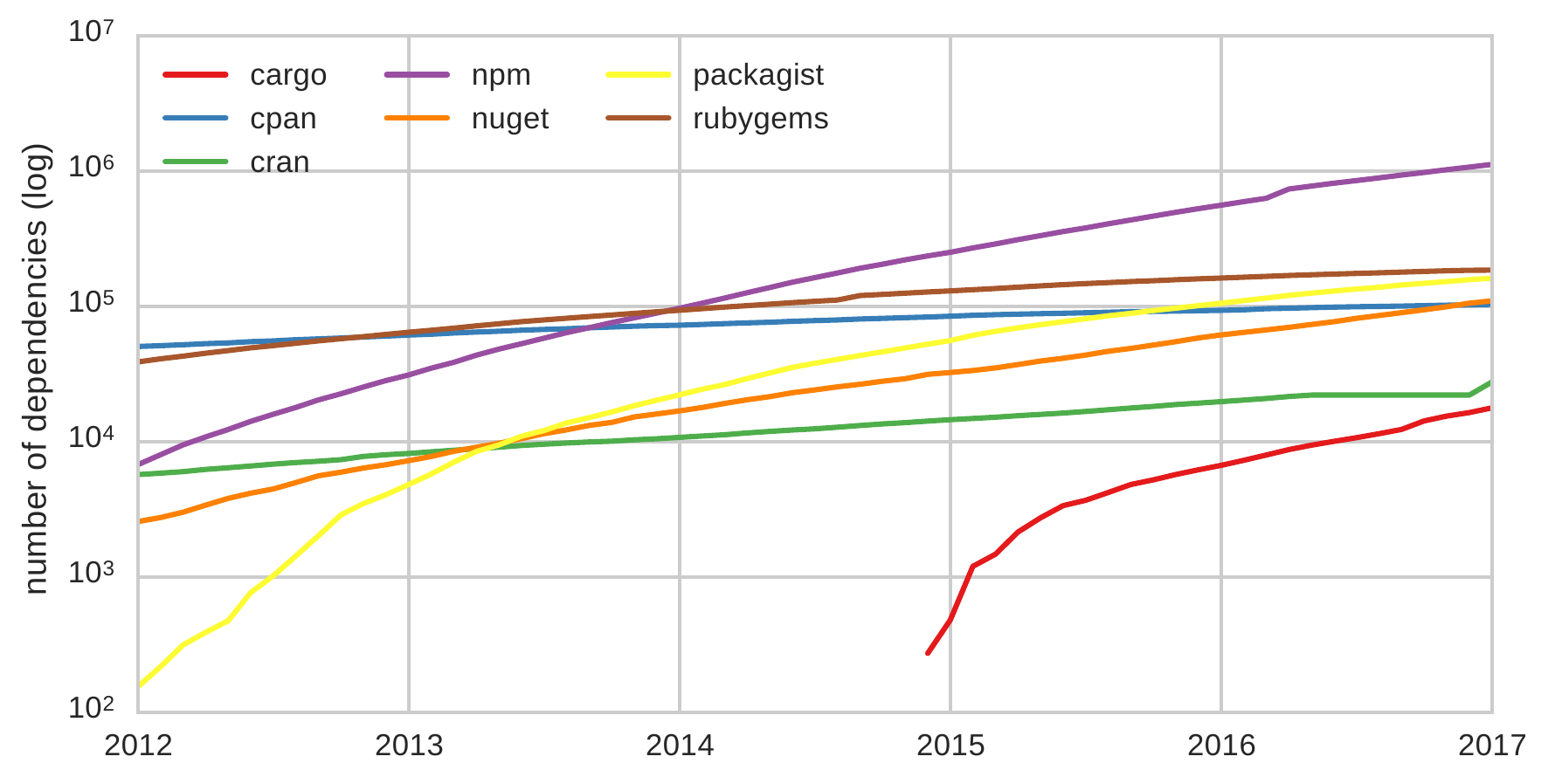}
  \caption{
  Evolution of the number of dependencies (considering for each point in time  the latest available release of each package).
  }
  \label{fig:graph_number_of_dependencies}   
\end{figure}

We computed the growth of each package dependency network by counting its number of nodes (packages) and edges (package dependencies).
The evolution of the number of packages is presented in \fig{fig:act_number_of_packages}, using a logarithmic scale for the y-axis. 
\fig{fig:graph_number_of_dependencies} presents the evolution of the number of dependencies for monthly snapshots of the dependency networks, and looks quite similar to the previous one.
We conclude that \textbf{all considered ecosystems continue to grow over time.}

\begin{table}[!htbp]
\caption{$R^2$-values of regression analysis on the evolution of the size metrics.}
\label{tab:growth_regressions}\centering
\begin{tabular}{l||c|c|c|c|c|c|c}
 %\hline
 \bf \# packages & \cargo & \cpan & \cran & \npm & \nuget & \packagist & \rubygems\\
  \hline%\hline
 linear & \bf 0.99 & \bf 0.97 & 0.84 & 0.83 & \bf 0.92 & 0.77 & 0.88 \\ %\hline
 exponential & 0.82 & 0.87 & \bf 0.97 & \bf 0.92 & 0.91 & \bf 0.89 & \bf 0.93 \\
 \multicolumn{1}{l}{} & \multicolumn{1}{l}{} & \multicolumn{1}{l}{} & \multicolumn{1}{l}{} & \multicolumn{1}{l}{} & \multicolumn{1}{l}{} & \multicolumn{1}{l}{} & \multicolumn{1}{l}{} \\
 
 % \bf releases & \cargo & \cpan & \cran & \npm & \nuget & \packagist & \rubygems\\
 %  \hline%\hline
 % linear & \bf 0.99 & \bf 0.91 & 0.88 & 0.80 & \bf 0.89 & 0.69 & 0.85 \\ %\hline
 % exponential & 0.72 & 0.88 & \bf 0.92 & \bf 0.93 & 0.88 & \bf 0.92 & \bf 0.93 \\
 % \multicolumn{1}{l}{} & \multicolumn{1}{l}{} & \multicolumn{1}{l}{} & \multicolumn{1}{l}{} & \multicolumn{1}{l}{} & \multicolumn{1}{l}{} & \multicolumn{1}{l}{} & \multicolumn{1}{l}{} \\

 \bf \# dependencies & \cargo & \cpan & \cran & \npm & \nuget & \packagist & \rubygems\\ 
 \hline%\hline
 linear & \bf 0.97 & \bf 1.00 & 0.98 & 0.83 & 0.89 & \bf 0.93 & \bf 1.00 \\ %\hline
 exponential & 0.85 & 0.98 & \bf 0.99 & \bf 0.98 & \bf 0.99 & 0.88 & 0.97 \\
\end{tabular}
\end{table}

To determine whether the dependency networks have a different speed of growth according to both size metrics, we carried out a regression analysis using different parametric growth models. The $R^2$ values reflecting the ``goodness of fit'' of the models\footnote{$R^2\in[0,1]$ and the closer to 1 the better the model fits the data.} are summarised in \tab{tab:growth_regressions}. Only the linear and exponential models are presented as these invariably have the highest $R^2$ values of all considered growth models.
We observe that \cargo and \cpan reveal a linear growth for both size metrics (with $R^2\geq 0.97$ in all cases). \cran and \npm are on the other side of the spectrum, with an observed exponential growth  for both size metrics (with $R^2\geq 0.92$ in all cases). \nuget falls somewhere in between, growing exponentially in number of dependencies, but linearly in number of packages. \packagist and \rubygems have the opposite behaviour, growing linearly in number of dependencies but exponentially in number of packages.

To find out if the number of dependencies is growing faster than the number of packages, we computed the ratio of the number of dependencies over the number of packages. While this ratio remains stable for \cpan, \packagist and \rubygems, it increases for \cargo, \cran, \npm and \nuget, suggesting an increasing complexity relative to the number of packages.

We conclude that \textbf{the increase in size and complexity varies across ecosystems}.
The observed differences do not seem to depend on the ecosystem size or age.
For example, \cran is one of the smallest and oldest ecosystems and \npm the largest and much more recent, but they both exhibit an exponential growth rate according to both size metrics.
We assume that external factors, such as the popularity of the ecosystem or the activity of its contributor community, play a role in this growth rate. Determining these external factors and how they influence the ecosystem growth remains a topic of future work.

\begin{mdframed}
\textbf{Summary.} To answer $RQ_1$ we studied the growth of package dependency networks over time, based on the number of packages and their dependencies.
We observed that the dependency networks of all studied ecosystems tend to grow over time, though the speed of growth may differ. We also analysed the ratio of dependencies over packages as a simple measure of the network's complexity, and observed that this complexity remains stable for some ecosystems, while it tends to increase for others.
\end{mdframed}

%%%%%%%%%%%%%%%%%%%%%%%%%%%%%%%%%%%%%%%%%%
\section{$RQ_2$: How frequently are packages updated?}
\label{sec:rq-change}

Updating a package to a new release, regardless of whether it contains new features, bug fixes or API changes, is a common and natural process for a maintainer. However, such package updates can often be quite challenging in presence of package dependencies~\cite{BenMorris2016}:
``\textit{Change in an API is inevitable as your knowledge and experience of a system improves. Managing the impact of this change can be quite a challenge when it threatens to break existing client integrations.}''
This threat is confirmed by previous empirical research observing that package updates may cause many maintainability issues or even failures in dependent packages~\cite{DiCosmo2008,Bavota2015,Bogart2016,Decan2017SANER}.

\begin{figure}[!htbp]
   \centering
   \includegraphics[width=\figsize]{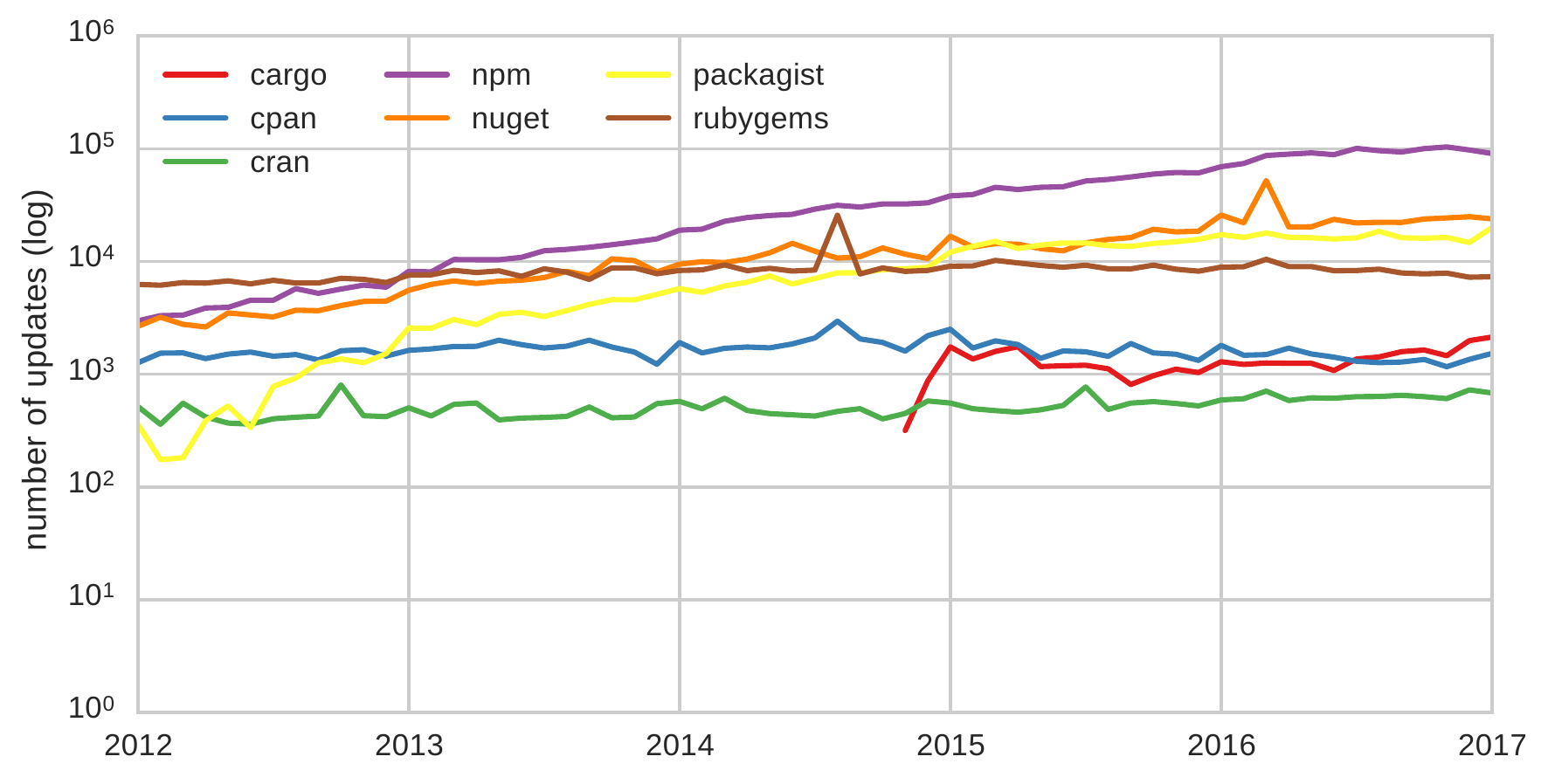}
   \caption{
     Evolution of number of package updates by month (using a logarithmic y-axis).
   }
   \label{fig:act_number_of_updates_by_month}
\end{figure}

To provide an upper bound estimate of how often such issues may arise, we compare across the considered ecosystems how frequently packages are being updated.
\fig{fig:act_number_of_updates_by_month} shows the evolution of the monthly number of package updates for each ecosystem.
We observe that, depending on the ecosystem, \textbf{the number of package updates either remains stable or tends to grow over time.}

For the smallest ecosystem \cargo and the two oldest ecosystems \cpan and \cran, the number of updates remains more or less stable. For \rubygems we observe a slight increase in the number of updates. For \npm, \nuget, and \packagist the observed growth is considerably larger.
We hypothesise that the frequency of package updates is related not only to the size of the ecosystem but also to the popularity of the ecosystem and its associated programming language.

The notable exception is \cran which, despite being linked to the popular \R language, exhibits a relatively low monthly number of updates. 
A plausible explanation is that \cran package maintainers are encouraged to limit the frequency of package updates 
because of \cran's ``rolling release'' policy that imposes packages to be up-to-date with their dependencies~\cite{CRAN}:
``\textit{Submitting updates should be done responsibly and with respect for the volunteers' time. Once a package is established (which may take several rounds), `no more than every 1--2 months' seems appropriate.}''

While \fig{fig:act_number_of_updates_by_month} provides a global view on the package update frequency, let us narrow this further down by distinguishing between \emph{required} and \emph{dependent} packages. Both types of packages face opposing forces influencing their update frequency.
On the one hand, required packages need to be updated regularly to take into account changes requested by the developers of their dependents.
On the other hand, dependent packages prefer to have limited updates of their dependencies as this may introduce backward incompatibilities. This is indeed a complaint of many package maintainers~\cite{Mens2015}: ``\emph{Especially with respect to package dependencies, the risk of things breaking at some point due to the fact that a version of a dependency has changed without you knowing about it is immense. That actually cost us weeks and months in a couple of professional projects I was part of.}''

\begin{figure}[!htbp]
   \centering
   \includegraphics[width=\figsize]{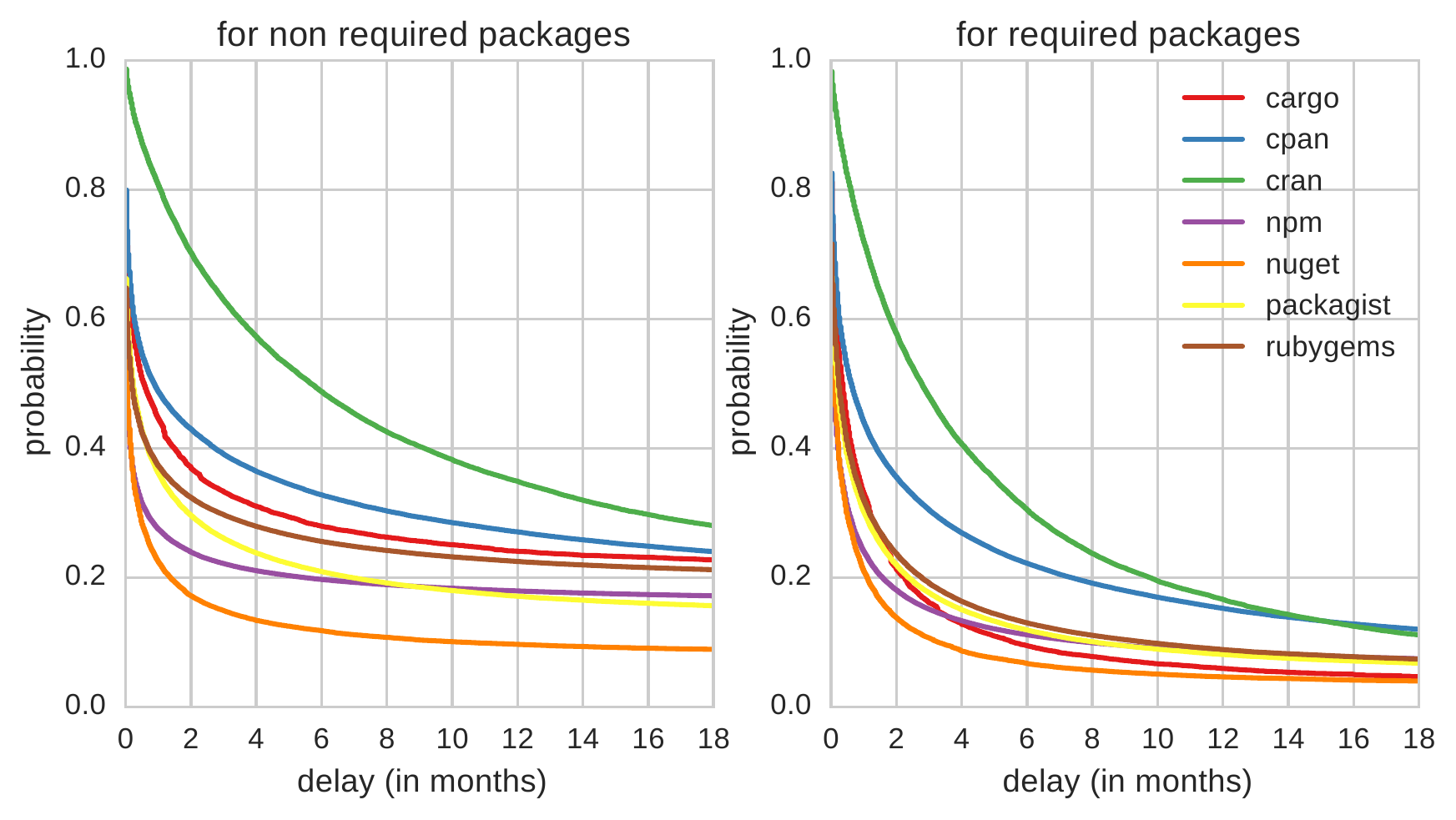}
   \caption{
     Survival probability of a package release (i.e., time until a more recent release becomes available). Left plot shows packages that are not required while right plot shows those that are required. 
   }
   \label{fig:act_survival_update_release}
\end{figure}

We therefore carry out a cross-ecosystem comparison of the time between successive releases of a package, by performing a survival analysis over the population of all package releases of each ecosystem. We distinguish between packages that are required and those that are not. For each release we consider the time required for a more recent release of the same package to become available in the package manager.
\fig{fig:act_survival_update_release} presents the Kaplan-Meier survival curves estimating the survival function of the probability that a release is not yet updated at time $t$.

Disregarding \cran, we observe a high similarity across ecosystems, with \textbf{a probability higher than 50\% for a package release to be updated within two months}, regardless of whether the package is required or not. The higher resilience of \cran packages to new updates can again be explained by \cran's policy, which is more demanding with respect to package updates.

We also observe that the population of required packages (\fig{fig:act_survival_update_release} right) receives updates considerably more frequently than those that are not required (\fig{fig:act_survival_update_release} left). Indeed, the survival curves for updates of required packages are invariably lower.
We statistically tested this observation by performing a log-rank test to compare the survival curves of non required packages to those of required packages. The test confirms with significance level $\alpha = 0.01$ that \textbf{required packages are updated significantly more often than packages that are not required}.

\begin{figure}[!htbp]
  \centering
  \includegraphics[width=\figsize]{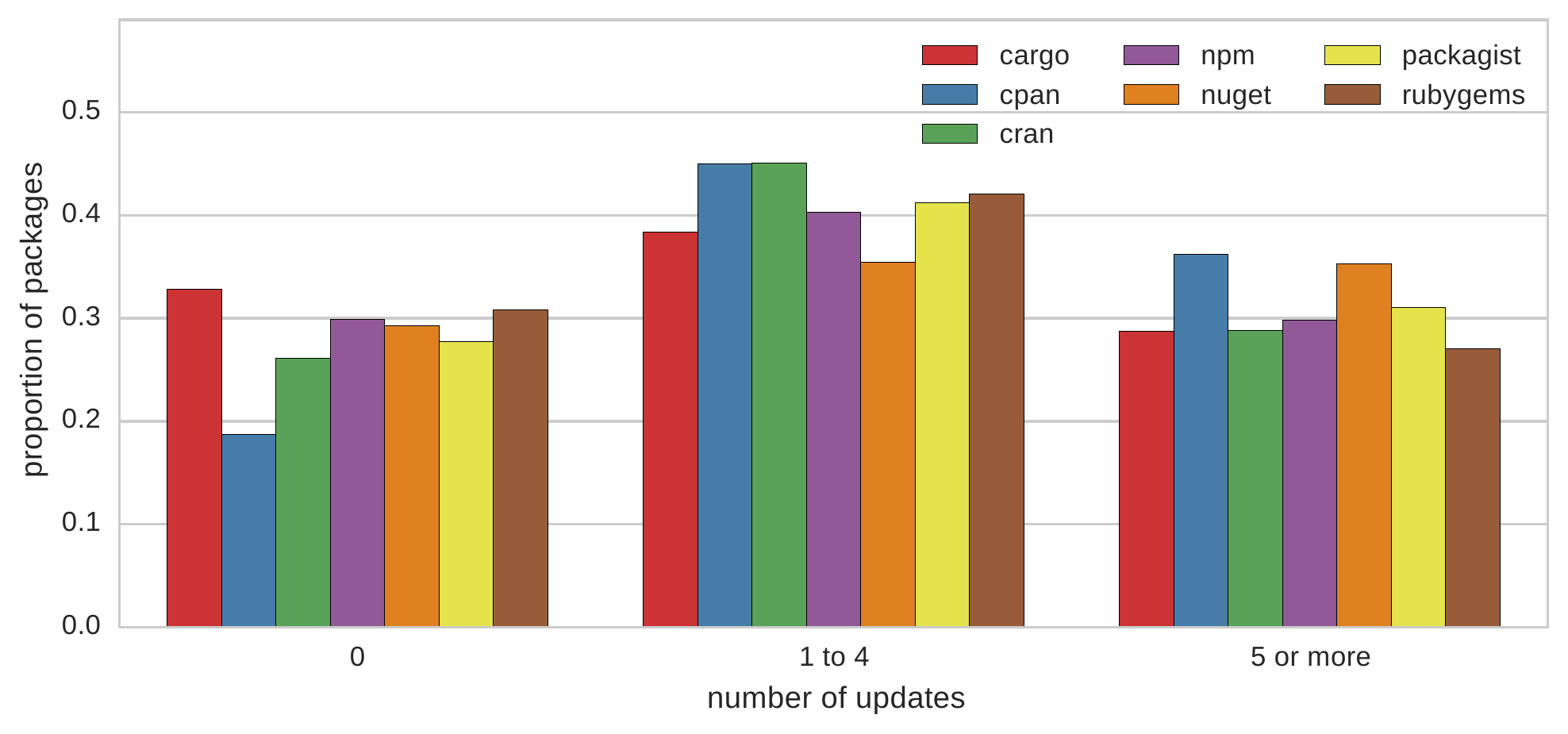}
  \caption{
   Proportion of packages having a given number of updates. 
  }
  \label{fig:act_nb_packages_updates_by_package}
\end{figure}

While the update frequency is rather similar for all ecosystems, let us drill even further down and consider the distribution of this frequency over individual packages.
\fig{fig:act_nb_packages_updates_by_package} compares the proportion of packages of each ecosystem having a given number of updates.
To facilitate visual comparison, we created three distinct bags corresponding to a more or less equal proportion (about one third each) of the total number of packages of the ecosystem.

With the exception of \cpan, between 26\% and 33\% of all packages were never updated,
between 35\% and 45\% of all packages were updated between 1 and 4 times, and between 27\% and 36\% of all packages were updated at least 5 times.
The higher proportion of updated packages for \cpan is arguably due to is age\footnote{\cpan is twice as old as the other considered ecosystems except for \cran.}: 
most of its packages were already available for years and, compared to the packages in the other ecosystems, had a significantly longer time to receive updates.

\begin{figure}[!htbp]
   \centering
   \includegraphics[width=\figsize]{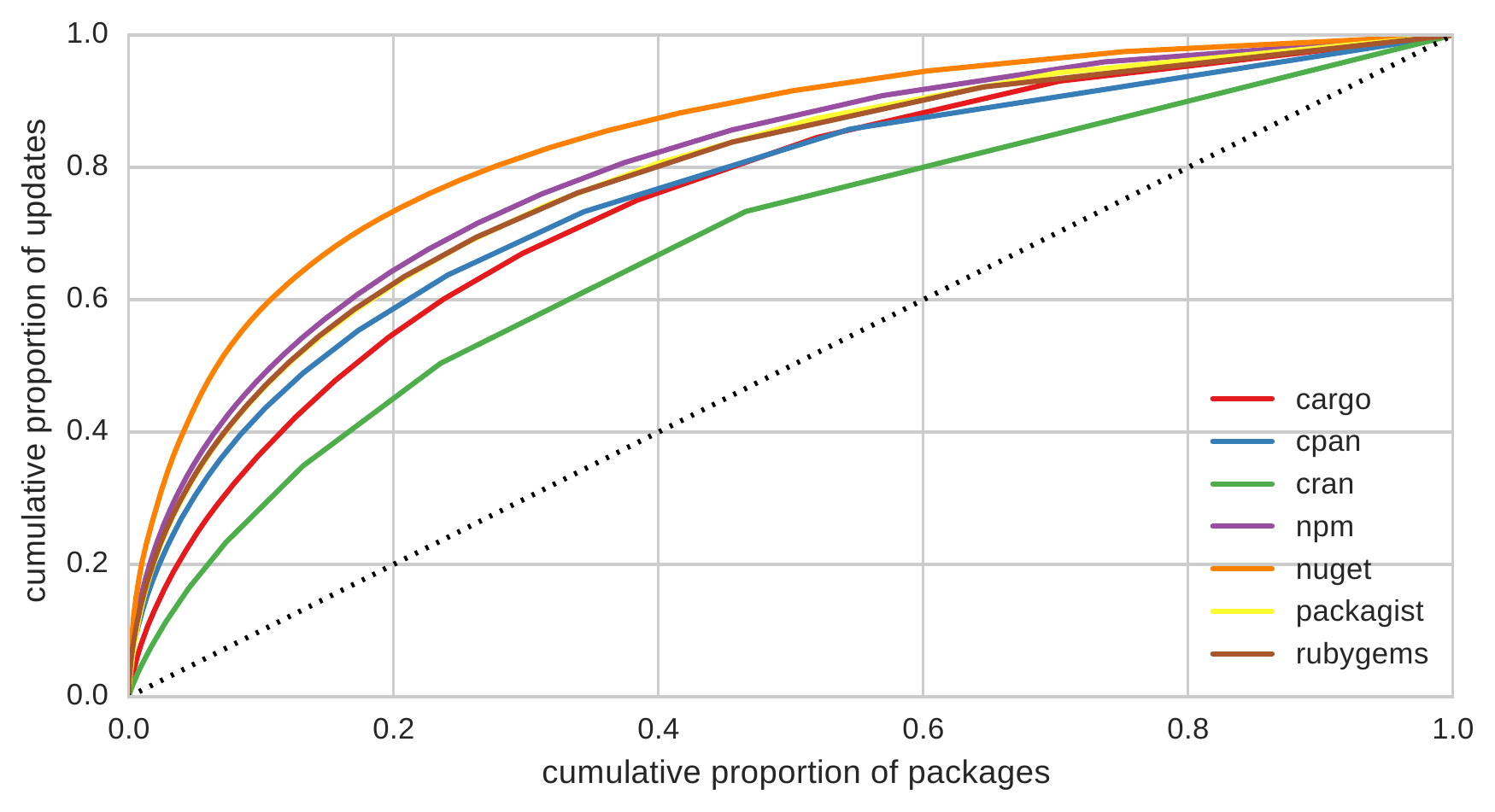}
   \caption{
     Inverted Lorenz curves for the number of package updates in 2016. Only packages that are at least updated once in 2016 are considered. 
   }
   \label{fig:act_lorenz_curve_2016}
% Proportion of packages required to reach at least 80% of all the updates:
% cargo: 0.4547304170905392
% cpan: 0.45109407144193003
% cran: 0.5992627698788836
% npm: 0.36405587651362037
% nuget: 0.2742631318952742
% packagist: 0.3902897706203316
% rubygems: 0.39703670942061037
\end{figure}

\fig{fig:act_nb_packages_updates_by_package} suggests that \textbf{the number of updates is not evenly distributed across packages}: regardless of the considered ecosystem, close to one third of all packages receive 5 or more updates, and close to one third of all packages receive no update at all.
\fig{fig:act_lorenz_curve_2016} presents an (inverted) Lorenz curve that sheds more light on the extent of the inequality in the distribution of the number of updates across packages.
It shows the cumulative proportion of updated packages responsible for the cumulative proportion of updates. 

To limit the statistical bias induced by packages that are not updated anymore, we only considered ``active'' packages that were updated at least once in 2016. These active packages represent between 15.9\% (for the oldest ecosystem \cpan) to 53.1\% (for the newest ecosystem \cargo) of the packages.
Note that, although \cran is the second oldest ecosystem in our list, its percentage of active packages is fairly high (34.4\%). This should not come as a surprise, since \cran's rolling release policy more or less forces packages to update regularly, to avoid them becoming archived.

Based on this figure, we do observe a difference in the distribution inequality for the different ecosystems. For all ecosystems except \cran, a minority of packages (from 27\% for \nuget to 45\% for \cargo) is responsible for more than 80\% of all package updates. In contrast, \cran has a more equal distribution: 60\% of all packages are required to reach 80\% of all package updates.

The inequality of these distributions can only partly be explained by the fact that required packages are updated more frequently (cf. \fig{fig:act_survival_update_release}). We therefore hypothesise that the package age (\ie the time elapsed since its first release was introduced) also plays an important role. The intuition is that younger (and hence, less mature) packages are more subject to changes than older (and hence, more stable) ones.

\begin{figure}[!htbp]
   \centering
   \includegraphics[width=\figsize]{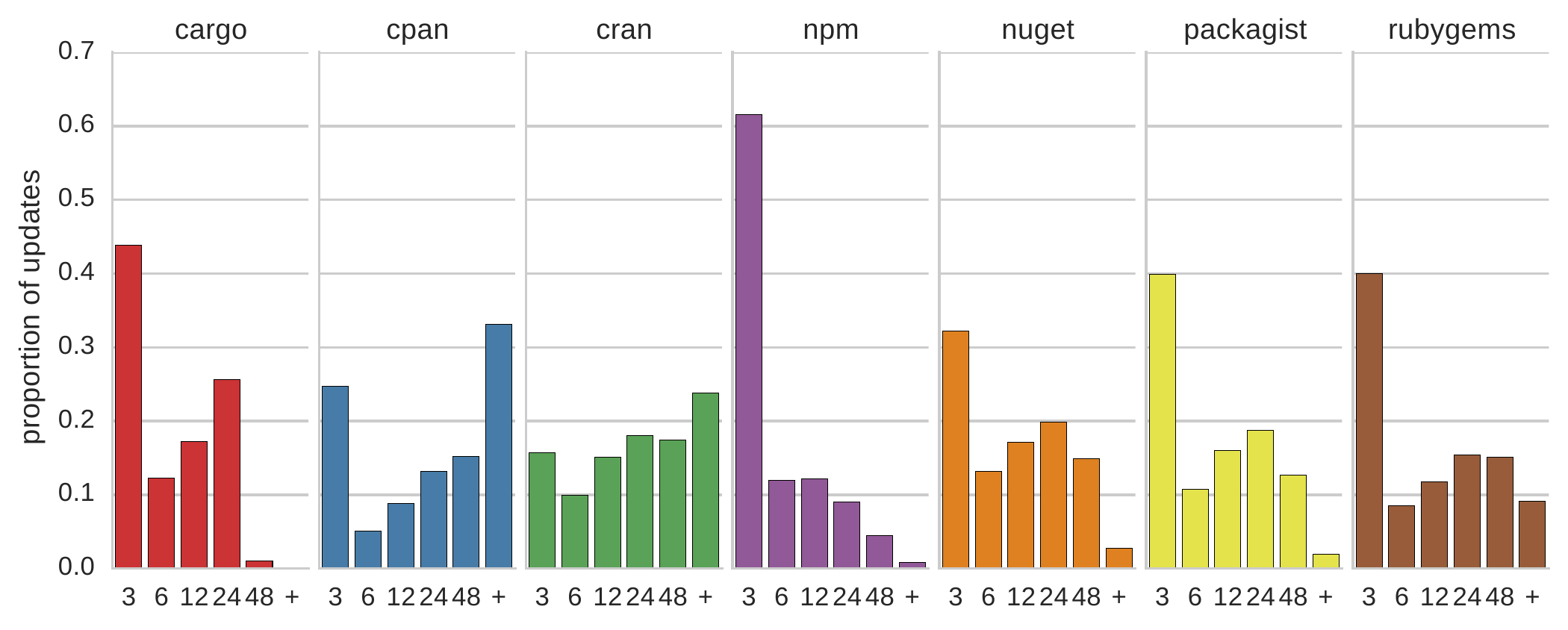}
   \caption{
     Proportion of updates in 2016 by package age (in months).
   }
   \label{fig:act_packages_updates_by_age_2016}
   % ecosystem 	cargo 	cpan 	cran 	npm 	nuget 	packagist 	rubygems
   % 3 	0.438302 	0.246777 	0.157141 	0.615736 	0.322516 	0.399165 	0.400375
   % 6 	0.123039 	0.050744 	0.099461 	0.119354 	0.131695 	0.107399 	0.085419
   % 12 	0.172254 	0.088189 	0.151228 	0.122018 	0.171262 	0.160109 	0.117631
   % 24 	0.256479 	0.131584 	0.179871 	0.090255 	0.198093 	0.187365 	0.154259
   % 48 	0.009926 	0.151648 	0.173959 	0.044821 	0.148649 	0.126673 	0.151230
   % + 	NaN 	0.331059 	0.238339 	0.007816 	0.027785 	0.019289 	0.091086
\end{figure}

\fig{fig:act_packages_updates_by_age_2016} presents the proportion of package updates in 2016 in terms of the age of the packages being updated. 
The results reflect our intuition. With the exception of \cpan and \cran, \textbf{the majority of the updates involve packages that are up to 12 months old.} 
For \cargo and \packagist, respectively 56\% and 50\% of the updates involve packages of less than 6 months. The inequality is even more pronounced for \npm, where more than 62\% of the updates are done for packages of less than 3 months old.

\cpan and \cran do not follow this rule. Indeed, the majority of package updates for them (respectively 58\% and 59\%) involve packages that are older than 2 years. We believe that this different change behavior is due to the fact that these two ecosystems are much older than the other ones.

\begin{mdframed}
\textbf{Summary.} We made the following observations in response to \emph{$RQ_2$: How frequently are packages updated?}
\begin{itemize}
\item The number of package updates in an ecosystem remains stable or tends to grow over time.
\item Most package releases are quickly updated within few months.
\item The number of package updates is distributed unequally: a minority of active packages is responsible for most of the package updates.
\item Young or required packages receive package updates more often.
\item Some of the observed behaviour appear to depend on the age of the ecosystem.
\end{itemize}
\end{mdframed}

\bigskip

Given that we have observed many similarities in the change dynamics of the considered ecosystems, but also some notable differences that appear to depend on the ecosystem's age, we wish to capture in a single time-dependent metric the specific characteristics that reflect the propensity for an ecosystem to change.
Such a metric must be comparable across ecosystems and must reflect both the amplitude and the importance of the considered ecosystem characteristics. 
Inspired by the famous Hirsch index~\cite{Hirsch2005}, which comprises in a single indicator a measure of both quantity and impact of the scientific output of a researcher~\cite{Costas2007193}, we therefore propose 
the following ecosystem \textit{Changeability Index}:

\begin{definition}
  The \textbf{Changeability Index} of an ecosystem $E$ at time $t$ is the maximal value $n$ such that there exist $n$ packages in $E$ at time $t$ having been updated at least $n$ times during the last month.\footnote{Because the choice of one month period may seem arbitrary, we also computed this index for several other periods, and did not observe different behaviours.} 
\end{definition}

By considering the $n$ most updated packages, this index takes into account the highly skewed distribution and the dispersion of updates we observed in Figures~\ref{fig:act_lorenz_curve_2016} and \ref{fig:act_packages_updates_by_age_2016}. It therefore appears to be an appropriate measure of both the amplitude (number of packages) and the importance (number of package updates) of the propensity for an ecosystem to change.
An important feature of this index is that it is largely independent of the ecosystem's size (expressed in number of packages). This makes it easy to compare the evolution of the index between ecosystems of varying sizes (cf. \tab{tab:characteristics}).

\begin{figure}[!htbp]
   \centering
   \includegraphics[width=\figsize]{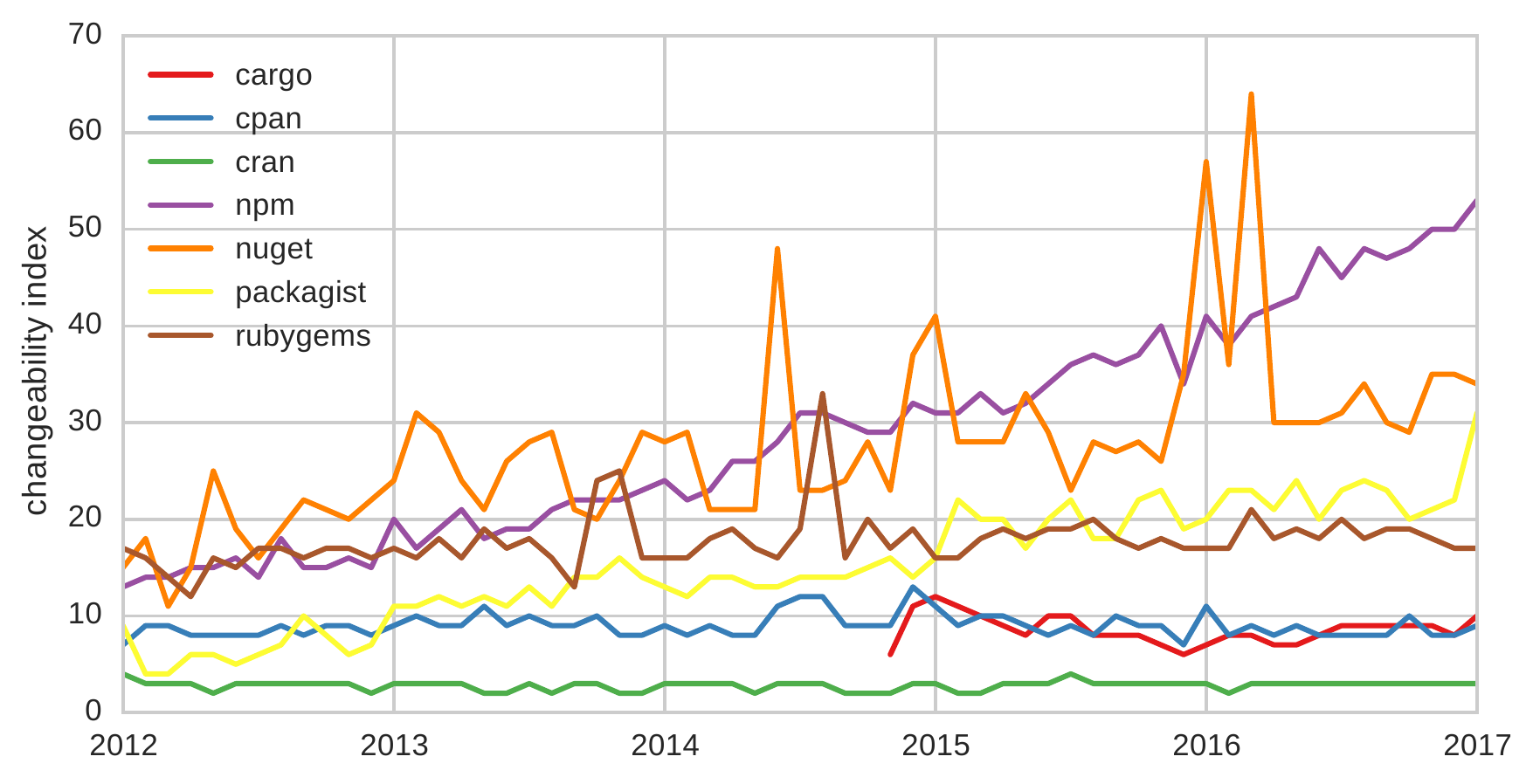}
   \caption{
     Cross-ecosystem comparison of the evolution of the Changeability Index.
   }
   \label{fig:act_volatility_index}
%  	cargo 	cpan 	cran 	npm 	nuget 	packagist 	rubygems
% min   6   7   2   13  11  4   12
% max   12  13  4   50  64  24  33
% last  10   9   3   53  34  31  17

% highest points for NuGet are due to TypeScript.DefinitelyTyped packages
% that were updated many times in a short period of time, coinciding to some
% important TypeScript releases. 
\end{figure}

\fig{fig:act_volatility_index} shows the evolution of the Changeability Index. 
Unsurprisingly, we find a low and constant value for \cran and \cpan, by far the two oldest ecosystems.
\cargo, \cpan and \rubygems also seem to have a more or less constant Changeability Index over time. 
\npm appears to be the most ``volatile'' ecosystem, reflected by the highest and fastest growing Changeability Index.
\nuget also features a high and increasing Changeability Index. For \nuget, we also observe some important peaks in June 2014 and early 2016, corresponding to a significant number of small, automatic and synchronised updates in a large number of packages related to the \textsf{TypeScript.DefinitelyTyped} project. These updates coincide with important releases of the \textsf{TypeScript} language of the Microsoft \dotnet platform.

%%%%%%%%%%%%%%%%%%%%%%%%%%%%%%%%%%%%%%%%%%
\section{$RQ_3$: To which extent do packages depend on other packages?}
\label{sec:rq-reuse}

One of the main reasons why packages depend on others is to enable software reuse, a basic principle of software engineering~\cite{Sametinger:1997:SER:260943}.
Dependencies allow packages to use the functionality offered by other packages (\eg libraries), avoiding the need to reimplement the same functionality.
Packaging ecosystems make it easier for developers to reuse code from other packages, by offering automated tools to manage multiple packages and their dependencies.
On the other hand, dependencies increase the risk of having important maintainability issues and failures~\cite{Bavota2015, Bogart2016}.
These failures can be caused by different events: a package may get removed entirely from the ecosystem, a package may become archived  because it no longer passes the quality checks or because its developer is no longer available, a package may be updated in backward incompatible ways, and so on.

Package maintainers share this concern. An Eclipse developer mentioned ``\textit{I only depend on things that are really worthwhile. Because basically everything that you depend on is going to give you pain every so often. And that's inevitable}''~\cite{Bogart2016}. A \cran developer stated 
``\textit{I had one case where my package heavily depended on another package and after a while that package was removed from \cran and stopped being maintained. So I had to remove one of the main features of my package. Now I try to minimize dependencies on packages that are not maintained by `established' maintainers or by me [...]}''~\cite{Mens2015}.
In earlier work, we observed that more than 40\% of the failures observed in \cran packages were caused by incompatible changes in required packages~\cite{Decan2016SANER}.

\begin{figure}[!htbp]
   \centering
   \includegraphics[width=\figsize]{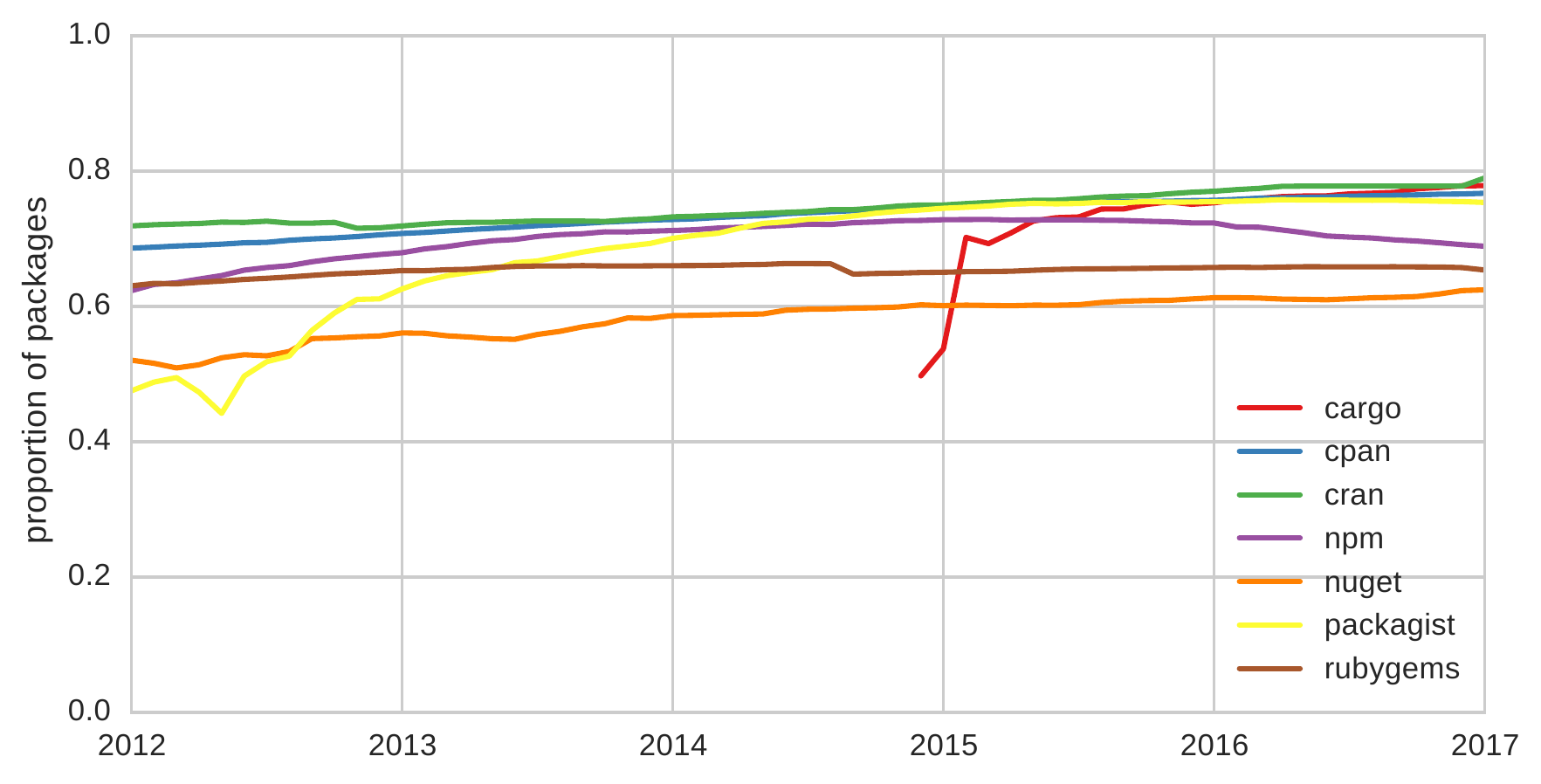}
   \caption{
     Proportion of connected packages.
   }
   \label{fig:graph_prop_connected_packages}
% min / max / latest
% cargo 	0.497585 	0.778736 	0.778736
% cpan 	0.686449 	0.767529 	0.767529
% cran 	0.715833 	0.790173 	0.790173
% npm 	0.623574 	0.728768 	0.689246
% nuget 	0.509383 	0.624806 	0.624806
% packagist 	0.442249 	0.758043 	0.753964
% rubygems 	0.630682 	0.663623 	0.654073

% proportion of packages in the biggest WCC
% min / max / latest, of all packages
% cargo 	0.299517 	0.753942 	0.753942
% cpan 	0.676429 	0.761203 	0.761203
% cran 	0.693741 	0.777708 	0.777708
% npm 	0.586312 	0.707404 	0.669457
% nuget 	0.371874 	0.553225 	0.553225
% packagist 	0.414894 	0.714125 	0.708849
% rubygems 	0.604599 	0.645605 	0.637845

% Proportion of connected packages in WCC
% cargo 	0.601942 	0.968161 	0.968161
% cpan 	0.985404 	0.991759 	0.991759
% cran 	0.969138 	0.984225 	0.984225
% npm 	0.940244 	0.970685 	0.97129
% nuget 	0.730049 	0.885435 	0.885435
% packagist 	0.938144 	0.942064 	0.940163
% rubygems 	0.958644 	0.972850 	0.97519
\end{figure}

Not all packages make use of dependencies in a similar way.
\fig{fig:graph_prop_connected_packages} shows the evolution over time of the proportion of connected packages, \ie packages that are either dependent or required. 
Regardless of the ecosystem, we observe that \textbf{a large majority of the packages are connected} (from 62\% for \nuget to 79\% for \cran in January 2017). 

Interestingly, smaller ecosystems (\cargo, \cpan, \cran and \packagist) exhibit a behaviour that is different from the larger ecosystems (\npm, \nuget and \rubygems). \textbf{Smaller ecosystems tend to have a higher proportion of connected packages, with an increasing trend over time.}

To verify if the connectedness of packages is spread over the entire ecosystem,
we computed the largest \emph{weakly connected component} for the latest snapshot of each dependency network. A weakly connected component of a directed graph is a subgraph in which each vertex is connected to every other vertex by an undirected edge path. 
We found that the overwhelming majority of connected packages (from 89\% for \nuget to 99\% for \cpan) are part of this component.

\begin{figure}[!htbp]
   \centering
   \includegraphics[width=\figsize]{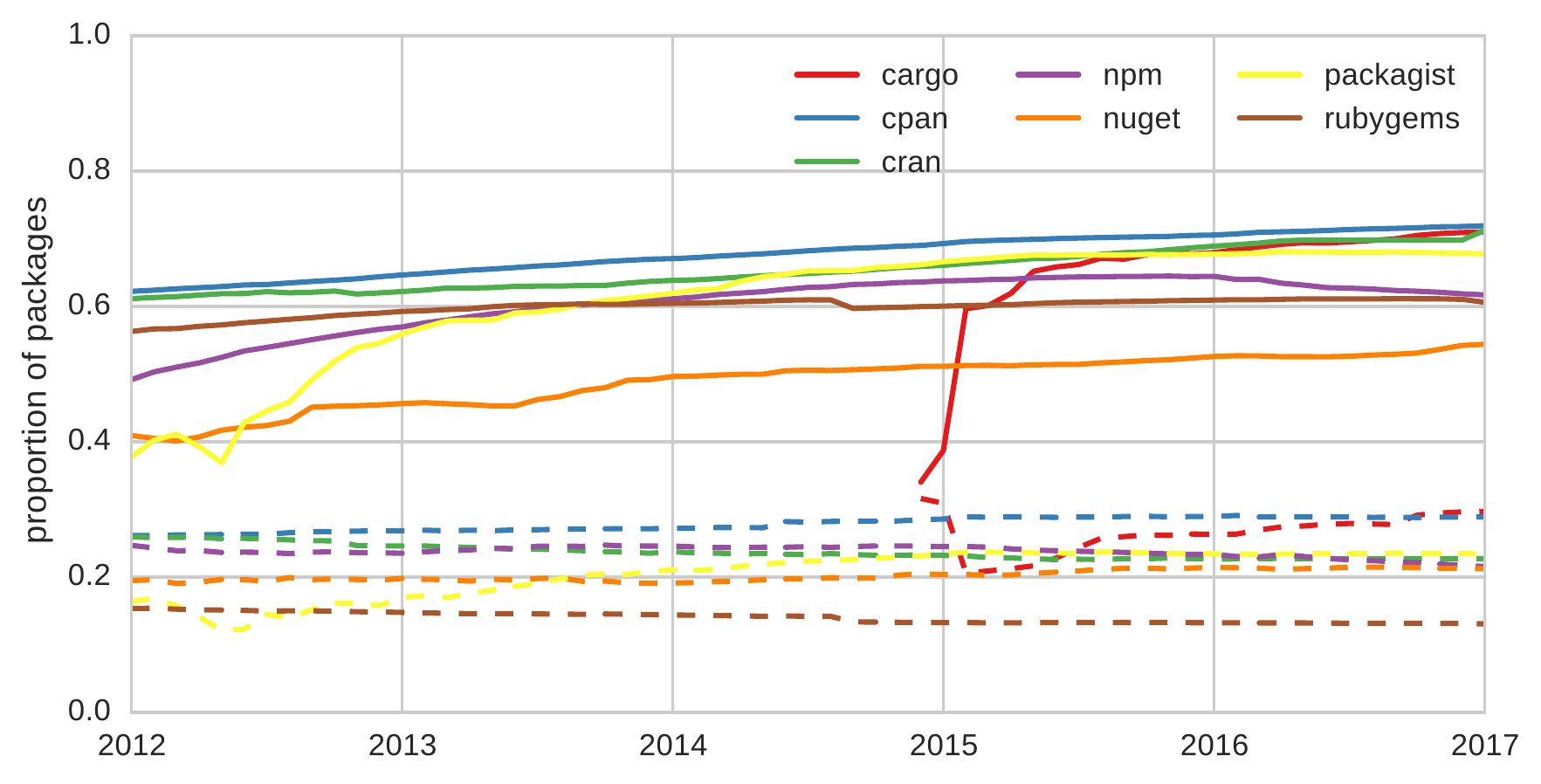}
   \caption{
     Proportion of dependent (straight lines) and required (dashed lines) packages.
   }
   \label{fig:graph_prop_required_dependent_packages}
% min / max / latest, first for required, then for dependent
% cargo 	0.206154 	0.316425 	0.297265
% cpan 	0.261604 	0.291340 	0.289431
% cran 	0.226225 	0.259726 	0.227351
% npm 	0.215789 	0.247508 	0.215789
% nuget 	0.190821 	0.214966 	0.212661
% packagist 	0.121581 	0.237397 	0.234486
% rubygems 	0.131153 	0.154178 	0.131153

% cargo 	0.340580 	0.710416 	0.710416
% cpan 	0.622569 	0.719538 	0.719538
% cran 	0.611811 	0.712673 	0.712673
% npm 	0.492015 	0.645115 	0.617236
% nuget 	0.401311 	0.544316 	0.544316
% packagist 	0.369301 	0.681301 	0.677855
% rubygems 	0.563313 	0.611619 	0.606441
\end{figure}

Given that a package can be connected either because it has dependents or because it requires packages (or both), we computed the proportion of dependent and required packages for each ecosystem. Their evolution is presented in
\fig{fig:graph_prop_required_dependent_packages}, and reveals that most packages are connected because they depend on other packages. 
We observe that \textbf{a majority of packages depends on a small minority of other packages} and that \textbf{the proportion of dependent packages  increases over time while the proportion of required packages remains quite stable}.

The fact that the behaviour of \cargo deviates from the other ecosystems in the beginning of \cargo's lifetime (end of 2014 -- early 2015) is very likely due to the fact that a larger proportion of packages was created to form the foundations or ``building blocks'' of the ecosystem on which future packages can rely. To a lesser extent, a similar behaviour can be observed for the \packagist ecosystem, that was created in 2012.

\begin{figure}[!htbp]
   \centering
   \includegraphics[width=\figsize]{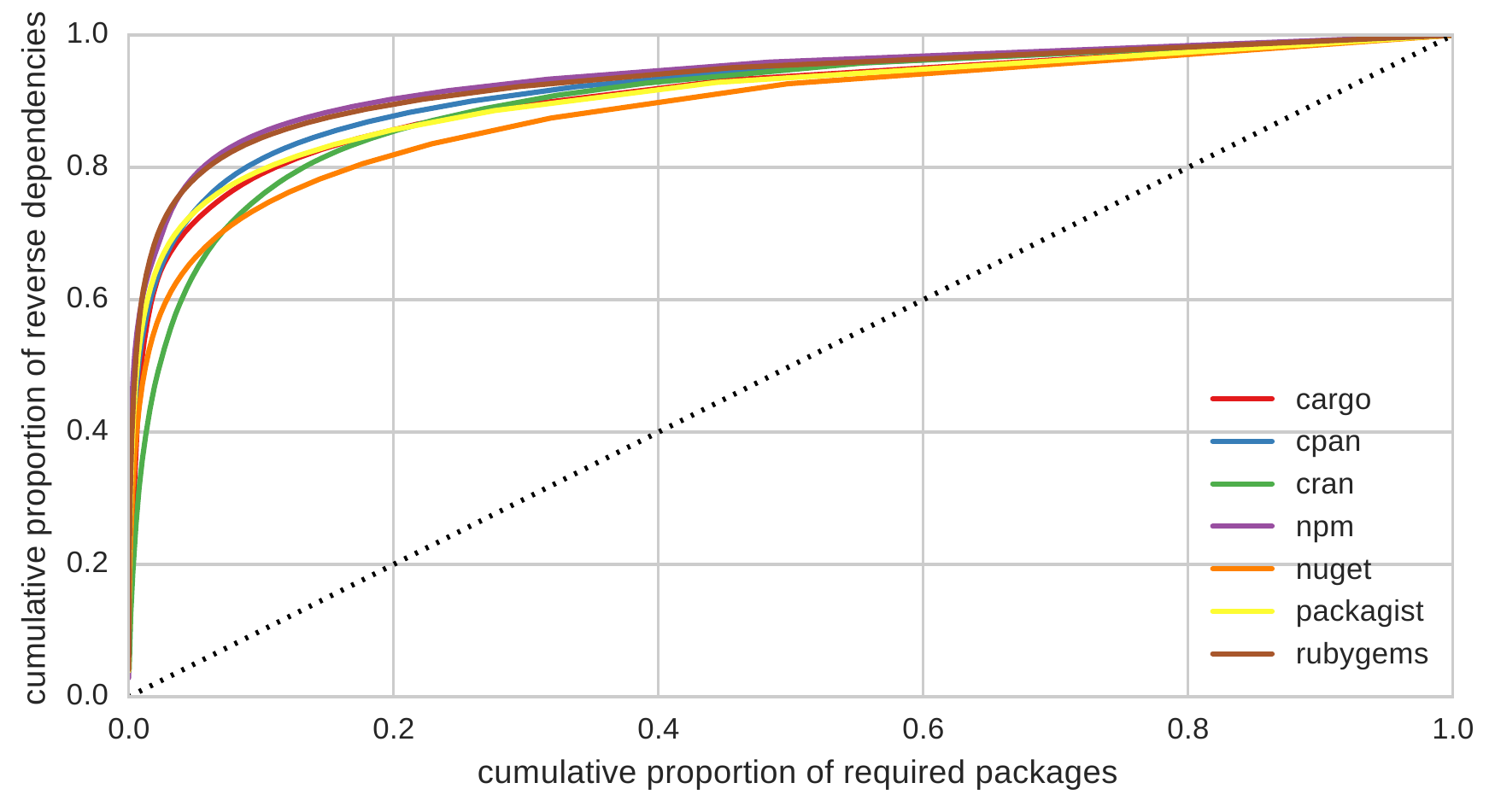}
   \caption{
     Inverted Lorenz curves for the number of reverse dependencies for required packages on 1 January 2017. 
   }
   \label{fig:graph_required_lorenz}
% 80%
% cargo: 0.11065759637188247
% cpan: 0.08856732348111825
% cran: 0.1319030591974568
% npm: 0.05590766335219416
% nuget: 0.16822719662676736
% packagist: 0.10421991801302066
% rubygems: 0.05960264900662213
\end{figure}

Not all required packages are equally required in terms of number of dependents. 
\fig{fig:graph_required_lorenz} shows an (inverted) Lorenz curve that represents the inequality among required packages, \ie the cumulative proportion of reverse dependencies in function of the cumulative proportion of required packages. 
We observe that \textbf{a very small proportion of required packages concentrates a very high proportion of reverse dependencies}. 
For instance, from only 6\% (for \npm and \rubygems) to 17\% (for \nuget) of required packages concentrate more than 80\% of all reverse dependencies.

\begin{figure}[!htbp]
   \centering
   \includegraphics[width=\figsize]{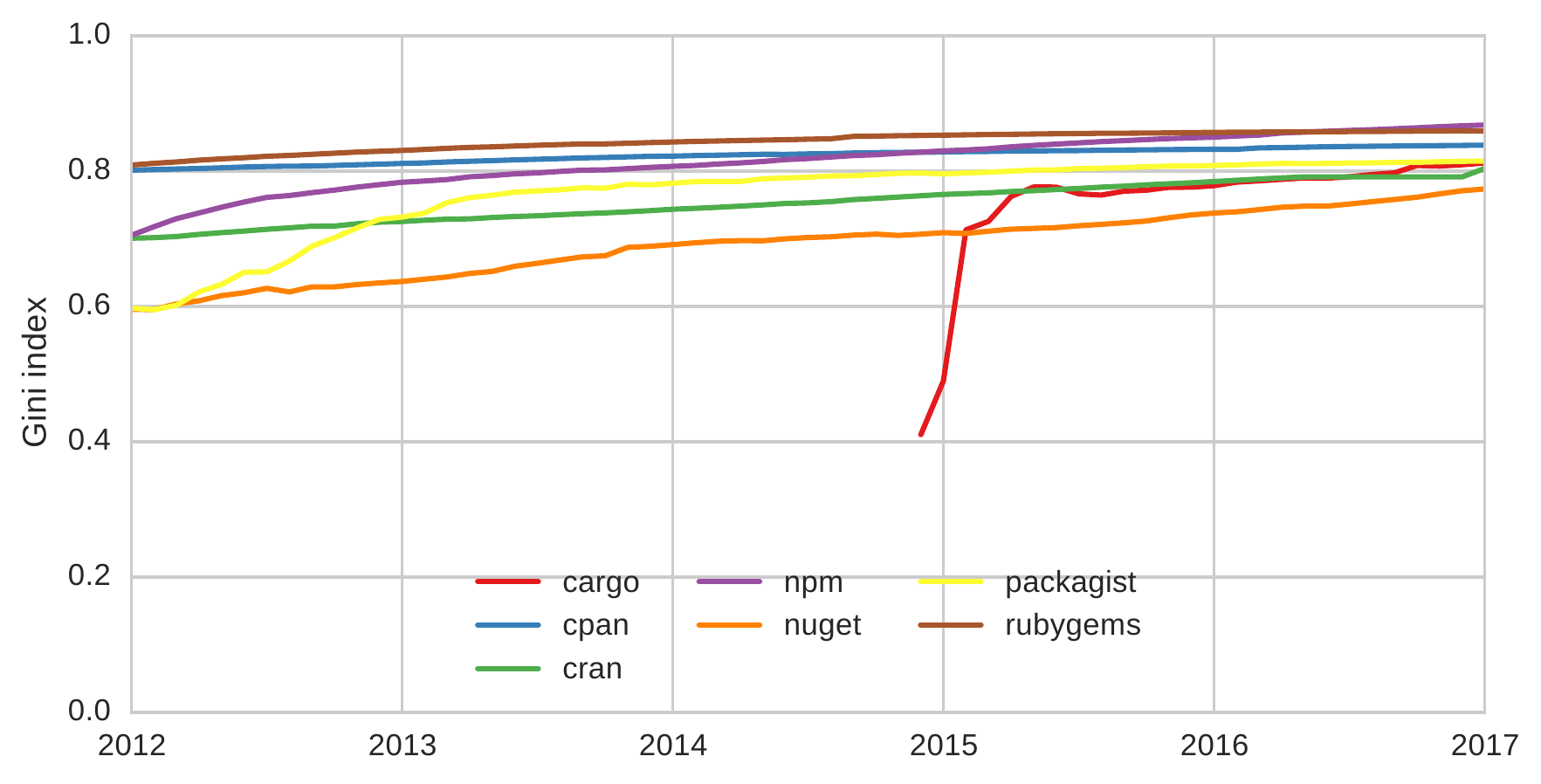}
   \caption{
     Evolution of the normalised Gini index reflecting the inequality of the number of dependent packages per required package.
   }
   \label{fig:graph_required_gini_index}
% min / max / latest
% cargo 	0.410780 	0.812388 	0.812388
% cpan 	0.801676 	0.838616 	0.838616
% cran 	0.701089 	0.804183 	0.804183
% npm 	0.705466 	0.868386 	0.868386
% nuget 	0.595607 	0.773817 	0.773817
% packagist 	0.595413 	0.815029 	0.815029
% rubygems 	0.809266 	0.859699 	0.859437
\end{figure}

To study how this unequal distribution changes over time, we computed the corresponding Gini inequality index for each month during the last five years. 
As the considered population of packages differ in size, and to allow comparisons accross ecosystems, we normalised the Gini index by dividing it by $1 - \frac{1}{n}$.
The results are shown in \fig{fig:graph_required_gini_index}.
We observe that the inequality index is similar and continuously increases for all ecosystems. 
On 1 January 2017, it ranges from 0.77 (for \nuget) to 0.87 (for \npm), indicating a very unequal distribution of the number of dependent packages in all ecosystems.

\begin{mdframed}[nobreak=true]
\textbf{Summary.} We made the following observations in response to \emph{$RQ_3$: To which extent do packages depend on other packages?}
\begin{itemize}
\item Dependencies are abundant in all packaging ecosystems.
\item Most packages are connected, mainly because they depend on other packages, and the proportion of connected packages increases over time.
\item Dependencies are not evenly spread across packages: less than 30\% of the packages are required by other packages, and less than 17\% of all required packages concentrate more than 80\% of all reverse dependencies. This unequal distribution of dependent packages increases over time.
\end{itemize}
\end{mdframed}

\bigskip

Similarly to how we characterised an ecosystem's propensity to change by means of a Changeability Index, we define an ecosystem's \emph{Reusability Index}. It comprises in a single indicator a measure of both the amplitude (number of required packages) and the extent (their number of dependent packages) of reuse. 
By considering the $n$ most required packages, this index takes into account the important inequality we observed in \fig{fig:graph_required_gini_index}.

\begin{definition}
  The \textbf{Reusability Index} of an ecosystem $E$ at time $t$ is the maximal value $n$ such that there exist $n$ required packages in $E$ at time $t$ having at least $n$ dependent packages. 
\end{definition}

\begin{figure}[!htbp]
   \centering
   \includegraphics[width=\figsize]{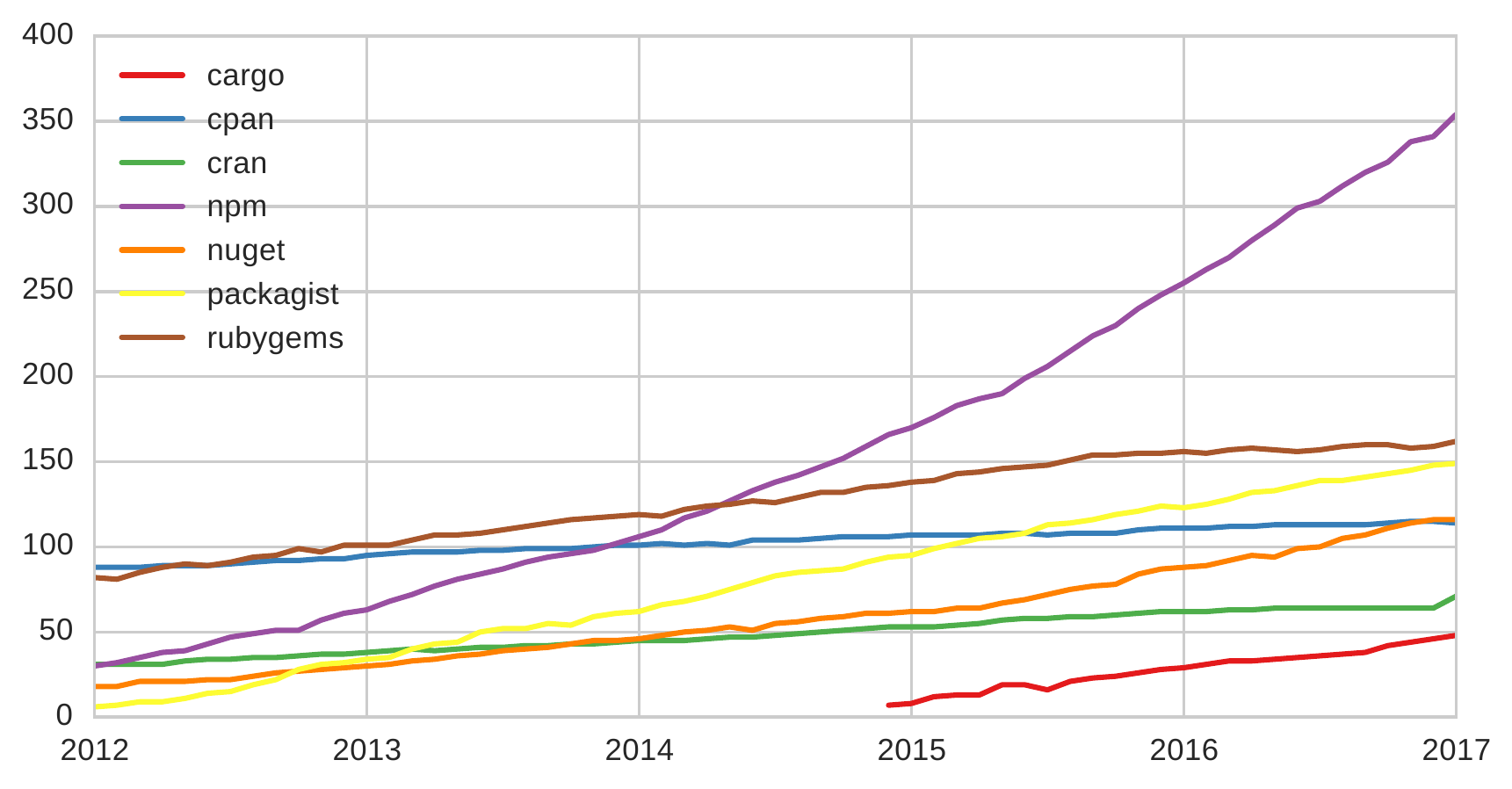}
   \caption{
     Cross-ecosystem comparison of the evolution of the Reusability Index. % direct dependent
   }
   \label{fig:graph_reusability_index}
   % Min / max / latest
  % cargo 	7.0 	48.0 	48
  % cpan 	88.0 	115.0 	114
  % cran 	31.0 	71.0 	71
  % npm 	30.0 	354.0 	354
  % nuget 	18.0 	116.0 	116
  % packagist 	6.0 	149.0 	149
  % rubygems 	81.0 	162.0 	162
\end{figure}

\fig{fig:graph_reusability_index} shows the evolution of the Reusability Index over time. 
We observe that it is increasing over time for all ecosystems, but at a different rate. We confirmed this through a regression analysis using different growth models. Both \npm and \nuget exhibit an exponential increase. The other ecosystems exhibit a linear increase, with a higher regression coefficient for \packagist, \cargo and \rubygems (respectively $0.08$, $0.05$ and $0.05$) than for the older ecosystems \cpan and \cran ($0.02$ for both). The obtained $R^2$ values are summarized in \tab{tab:reuse_regressions}.

\begin{table}[!htbp]
\caption{$R^2$-values of regression analysis on the evolution of the Reusability Index.}
\label{tab:reuse_regressions}\centering
\begin{tabular}{l||c|c|c|c|c|c|c}
 %\hline
 \bf  & \cargo & \cpan & \cran & \npm & \nuget & \packagist & \rubygems\\
  \hline%\hline
 linear & \bf 0.98 & \bf 0.98 & \bf 0.99 & 0.97 & 0.97 & \bf 1.00 & \bf 0.98\\ %\hline
%\alex{Should I include the regression coefficients? It is of respectively 0.05, 0.02, 0.02, 0.18, 0.05, 0.08, 0.05}
% coefficient of linear regression & 0.05 & 0.02 & 0.02 & 0.18 & 0.05 & 0.08 & 0.05\\
 exponential & 0.90 & 0.97 & 0.98 & \bf 0.98 & \bf 0.99 & 0.85 & 0.97
\end{tabular}
\end{table}

The higher values and growth rate for \npm could be explained by the relative poorness of \javascript's standard library. Unlike the standard libraries of most other languages, the one of \javascript is kept intentionally small for reasons explained by its creator Brendan Eich~\cite{Hemel2010}:
``\textit{The real standard library people want is more like what you find in \python or \ruby, and it's more batteries included, feature complete, and that is not in \javascript. That's in the \npm world or the larger world.}''
The result of this is that \npm contains a large and increasing number of packages that provide basic functionality on which many other packages depend.

%%%%%%%%%%%%%%%%%%%%%%%%%%%%%%%%%%%%%%%%%%
\section{$RQ_4$: How prevalent are transitive dependencies?}\label{sec:rq-complexity}

While $RQ_3$ %\sect{sec:rq-reuse}
focused on the presence of \emph{direct} dependencies between packages, $RQ_4$ focuses on the additional ``hidden'' reuse induced by \emph{transitive} dependencies. 
Transitive dependencies may cause package failures to potentially affect many other packages.
Such highly transitively required packages represent a potential Achilles' heel in an ecosystem: breaking or removing only one of them can impact a large proportion of the other packages in the ecosystem.

A striking example of this was experienced in \npm in March 2016. The sudden and unexpected removal of a package called \textsf{left-pad} had a large impact on the ecosystem, breaking over five thousand transitive dependents ($> 2\%$ of all \npm packages at that time), including packages whose maintainers were not even aware they depend on it:
``\textit{This impacted many thousands of projects. [...] We began observing hundreds of failures per minute, as dependent projects -- and their dependents, and their dependents... -- all failed when requesting the now-unpublished package.}''~\cite{NPM2016}

As another example, in November 2010, release 0.5.0 of \textsf{i18n} in \rubygems notably broke the popular \textsf{ActiveRecord} package, on which relied over 900 packages ($> 5\%$ of all packages).

\begin{figure}[!htbp]
   \centering
   \includegraphics[width=\figsize]{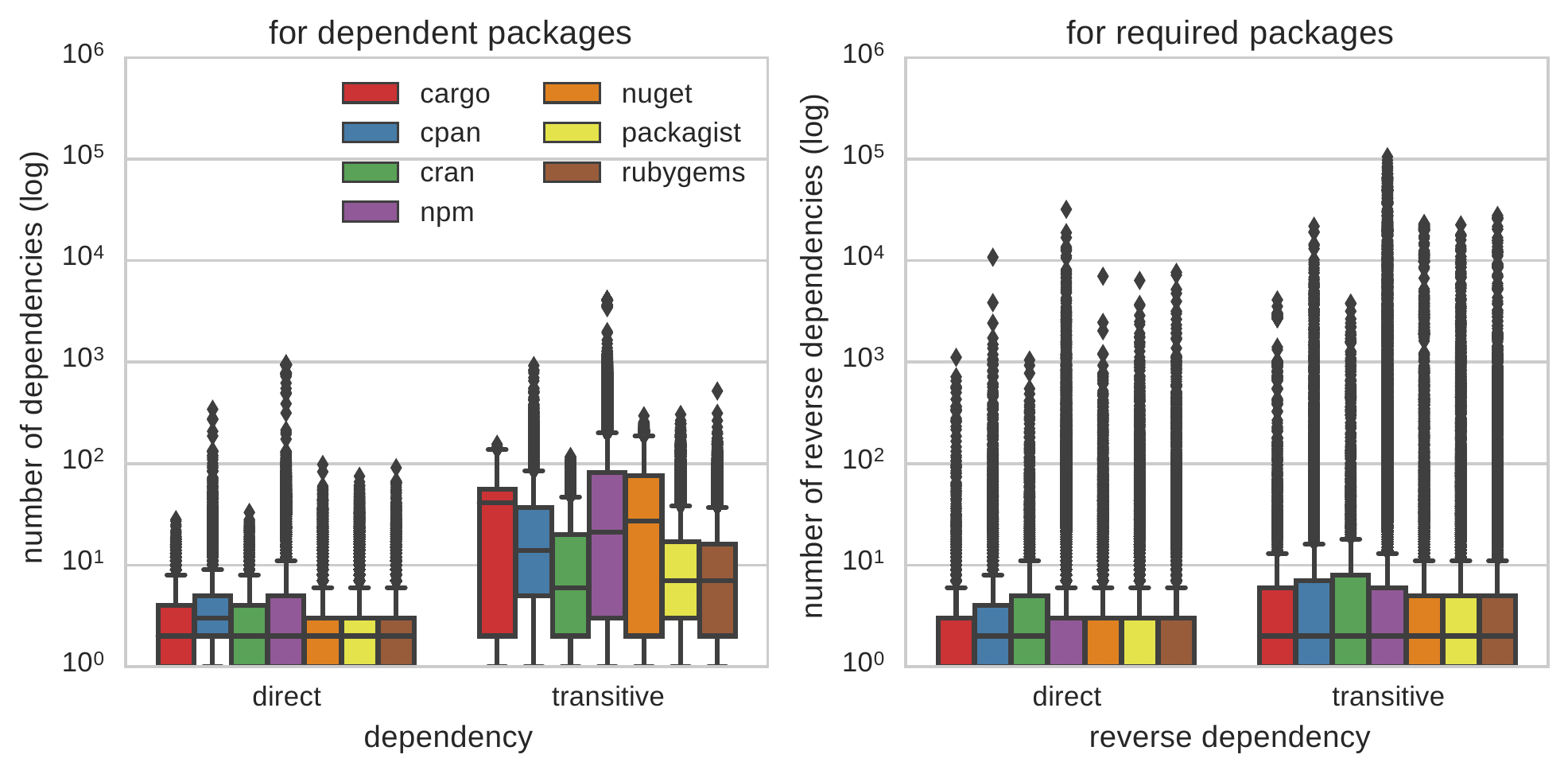}
   \caption{
       Distribution of the number of dependencies for dependent packages (left) and of reverse dependencies for required packages (right), in January 2017.
     }
   \label{fig:graph_distribution_dependencies_non_isolated}
\end{figure}

To reveal the prevalence of transitive dependencies in the studied ecosystems, the boxplots in \fig{fig:graph_distribution_dependencies_non_isolated} show the distribution of the number of direct and transitive dependencies for dependent packages (left), and reverse direct and reverse transitive dependencies for required packages (right) for comparison.
We observe that, \textbf{while a majority of the dependent packages have few direct dependencies, they have a much higher number of transitive dependencies}.
For instance, half of the dependent packages in \cargo, \npm and \nuget have at least 41, 21 and 27 transitive dependencies, respectively, where their median number of direct dependencies is only 2. 

\begin{figure}[!htbp]
   \centering
   \includegraphics[width=\figsize]{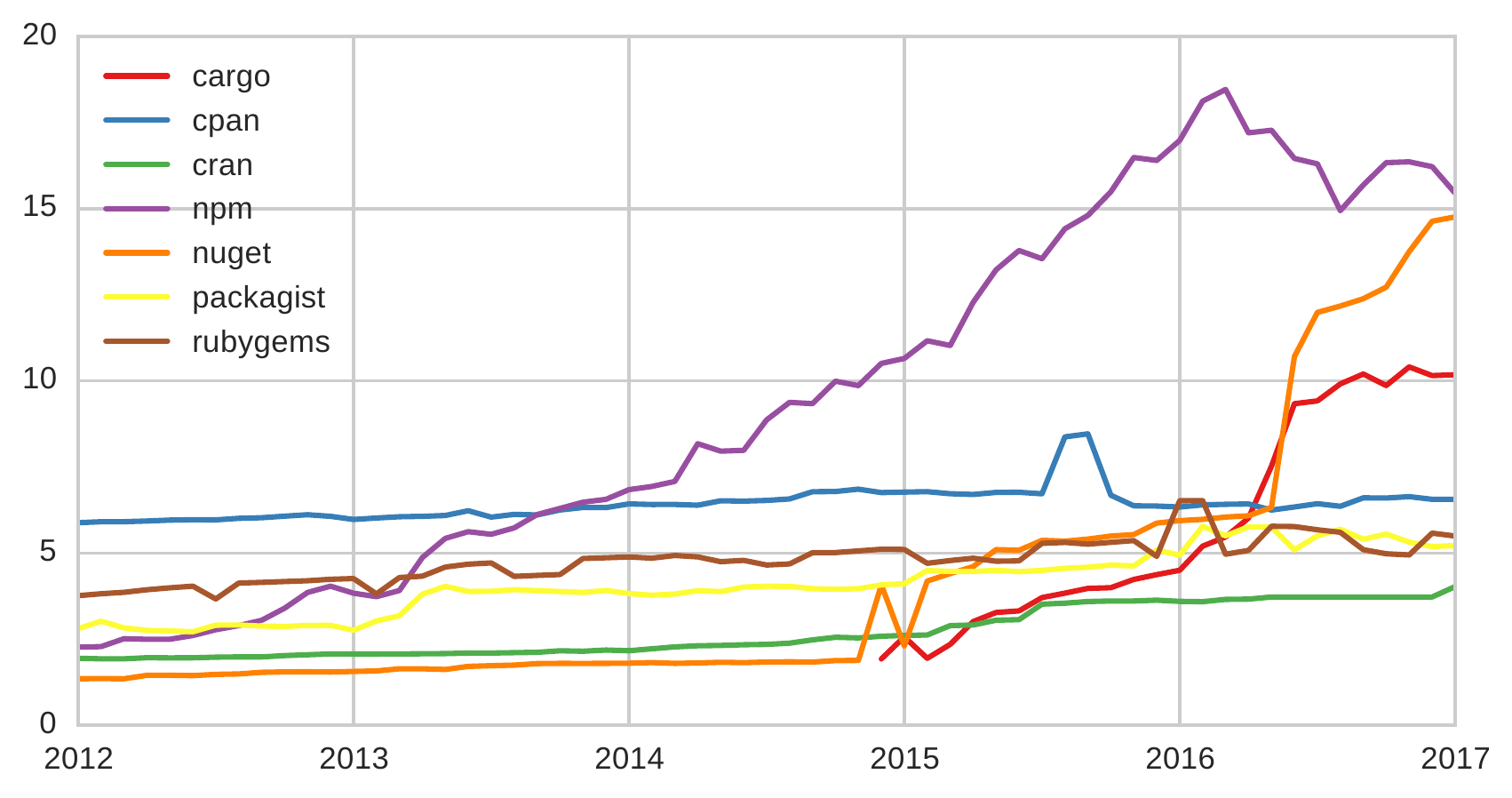}
   \caption{
     Evolution of the ratio between the number of transitive dependencies and the number of direct dependencies. 
   }
   \label{fig:graph_ratio_transitive_per_direct}
% min / max / latest
% cargo 	1.927007 	10.407472 	10.1804
% cpan 	5.883760 	8.464031 	6.56367
% cran 	1.930371 	4.028118 	4.02812
% npm 	2.269474 	18.464631 	15.4561
% nuget 	1.351852 	14.764984 	14.765
% packagist 	2.716883 	5.778226 	5.2198
% rubygems 	3.668201 	6.525641 	5.49659   
\end{figure}

\fig{fig:graph_ratio_transitive_per_direct} shows the evolution of the ratio between the total number of transitive dependencies and the total number of direct dependencies.
For \cpan, \cran, \packagist and \rubygems this ratio is stable over time, while it is increasing for the three other ecosystems.
In January 2017, it is even from 2 to 3 times higher for \cargo, \npm and \nuget than for the other ecosystems.
The observed peak for \cpan in July/August 2015 is the result of a temporary change in the list of dependencies of package \textsf{ExtUtils-MakeMaker}. During those two months, this highly required package (with more than 16k transitive dependents) transitively relied on 11 additional packages, leading each of those transitive dependents to have 11 additional transitive dependencies.

The observed significant variations starting from early 2016 can be explained by local phenomena. 
For \npm, the decrease of the ratio is most likely a reaction to the aforementioned \textsf{left-pad} incident.
For \cargo, the observed increase was caused by the appearance of around 500 additional dependents for a set of strongly connected packages with many dependencies, including among others the popular \textsf{clippy}, \textsf{quickcheck}, \textsf{regex}, \textsf{simd} and \textsf{serde} packages.
For \nuget, we identified that \textsf{Newtonsoft.Json}, a package with 30 transitive dependencies, gained in few months more than 1,700 (resp. 2,100) additional dependents (resp. indirect dependents).

While maintainers are usually aware of the direct dependencies of their packages because they explicitly declare them,
they typically have a much less clear idea on which packages they depend indirectly, because most tools that help developers in managing dependencies do not take transitive dependencies into account, even though such transitive dependencies can be very numerous.

For example, on 1 January 2017, a package such as the popular \textsf{react} in \npm has only 3 direct dependencies, but transitively depends on 12 additional packages.
As a consequence, each of the 7,296 packages that directly depends on  \textsf{react} implicitly requires 15 additional packages.

Not only does the number of indirect dependencies contribute to the difficulty of identifying required packages, but also because these dependencies can be deeply nested in the dependency tree. 
Consider \textsf{co}, one of the most required packages in \npm. This package has 2,507 direct dependents and 51,497 indirect dependents. More than 50\% of its indirect dependents require \textsf{co} at a depth $\ge 5$, \ie it is a dependency of a dependency of a dependency of an indirect dependency.

\begin{figure}[!htbp]
   \centering
   \includegraphics[width=\figsize]{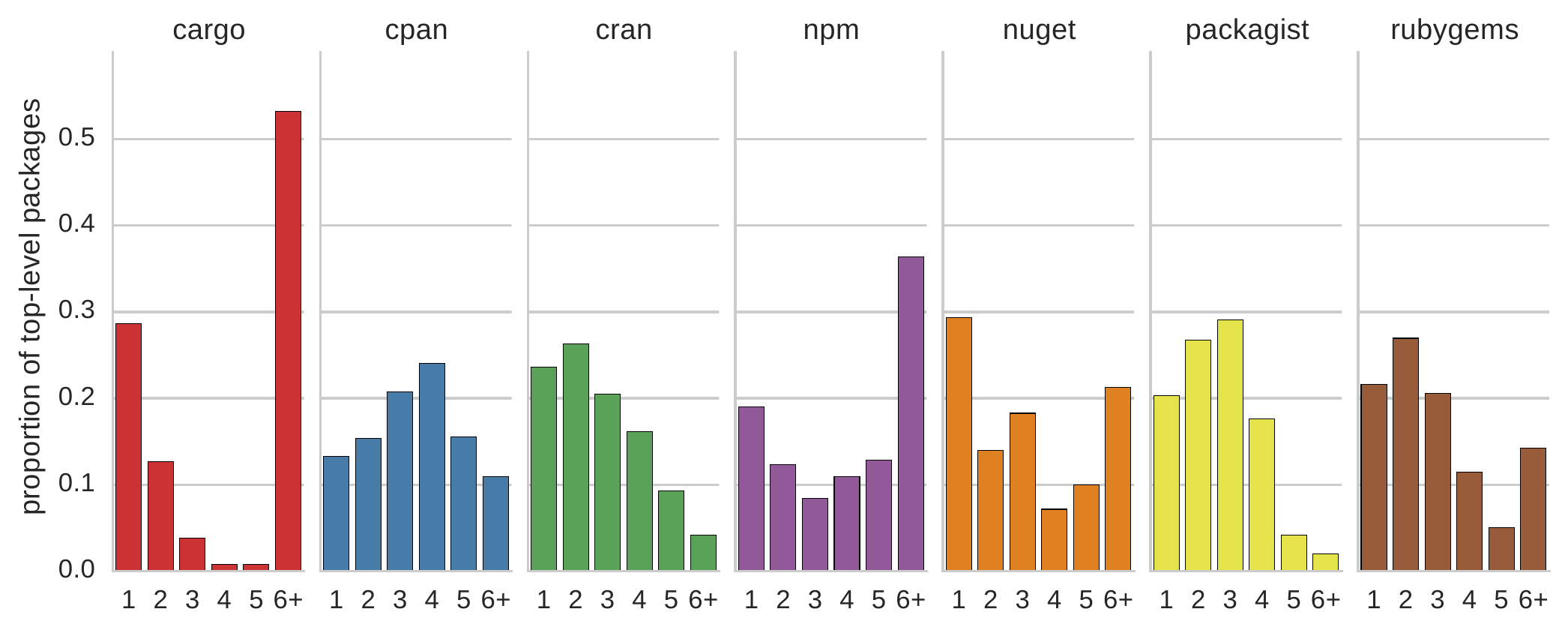}
   \caption{
     Proportion of top-level packages by depth of their dependency tree, in January 2017.
   }
   \label{fig:graph_toplevel_proportion_exact_depth}
% variable 	1 	2 	3 	4 	5 	6 or more
% cargo 	0.286794 	0.126749 	0.038332 	0.008114 	0.007834 	0.532177
% cpan 	0.133284 	0.153717 	0.207627 	0.240171 	0.155705 	0.109496
% cran 	0.236238 	0.263040 	0.204622 	0.161451 	0.093083 	0.041566
% npm 	0.190174 	0.123707 	0.084293 	0.109782 	0.128692 	0.363353
% nuget 	0.293529 	0.139555 	0.182780 	0.071744 	0.099655 	0.212737
% packagist 	0.203377 	0.267523 	0.291025 	0.176077 	0.042142 	0.019856
% rubygems 	0.216493 	0.269530 	0.205707 	0.114599 	0.050895 	0.142776

% CUMULATIVE:
% variable 	1 	2 	3 	4 	5 	6 or more
% cargo 	1.0 	0.713206 	0.586458 	0.548125 	0.540011 	0.532177
% cpan 	1.0 	0.866716 	0.712999 	0.505372 	0.265201 	0.109496
% cran 	1.0 	0.763762 	0.500722 	0.296100 	0.134649 	0.041566
% npm 	1.0 	0.809826 	0.686120 	0.601827 	0.492044 	0.363353
% nuget 	1.0 	0.706471 	0.566916 	0.384136 	0.312392 	0.212737
% packagist 	1.0 	0.796623 	0.529100 	0.238076 	0.061998 	0.019856
% rubygems 	1.0 	0.783507 	0.513977 	0.308270 	0.193671 	0.142776   
\end{figure}

% Proportion of top-level packages
% min / max / latest
% cargo 	0.181159 	0.509674 	0.481606
% cpan 	0.424627 	0.478158 	0.478158
% cran 	0.459558 	0.562822 	0.562822
% npm 	0.376426 	0.491136 	0.473606
% nuget 	0.318562 	0.412171 	0.412171
% packagist 	0.311111 	0.523901 	0.519749
% rubygems 	0.477878 	0.527434 	0.523403

To illustrate that \textsf{co} is not an isolated case, we computed the depth at which transitively required packages can be found. 
For this purpose, we consider \emph{top-level packages}, \ie packages that depend on other packages but that are not required themselves. 
Such top-level packages hence constitute the periphery of the dependency network, and their transitive closure will cover all dependencies of all packages.
Top-level packages represent between 41\% and 56\% of all the packages available in January 2017. 

\fig{fig:graph_toplevel_proportion_exact_depth} shows the proportion of top-level packages having a dependency tree of given depth in January 2017. 
\textbf{Regardless of the ecosystem, the majority of top-level packages have a deep dependency tree.} More than half of the top-level packages have a dependency tree depth of at least $3$. 

Some ecosystems have an even deeper nesting of dependencies. 
For \npm, more than 50\% of its top-level packages have a dependency tree depth of at least $5$.
We hypothesise that this is a combination of the recent surge in popularity of the ecosystem, combined with the lack of an extensive standard library, leading developers to rely on other packages even for basic features. 

Similarly, more than 50\% of the top-level \cargo packages have a dependency tree depth of at least $6$, and 25\% of the top-level \cargo packages have a dependency tree depth of at least $10$.
We assume that this is mainly related to the very young age of the \rust language and its \cargo package manager, leading developers to first try to develop smaller building bricks that are only a thin layer over previous ones and that can then be used by other packages to provide more ``high-level'' libraries such as those available in other languages.

\begin{mdframed}
\textbf{Summary.} We made the following observations in response to \emph{$RQ_4$: How prevalent are transitive dependencies?}
\begin{itemize}
\item For each ecosystem, the majority of dependent packages have few direct dependencies but a high number of transitive dependencies.
\item More than half of the top-level packages have a dependency tree of depth 3 or higher.
\end{itemize}
\end{mdframed}

\bigskip

Given that dependencies cause package failures to propagate to its dependents, and given the prevalence of transitive dependencies in each ecosystem, we are interested in a metric that reflects the fragility of an ecosystem because of the presence of highly required packages that may impact large parts of the ecosystem. We propose the parametric \emph{P-Impact Index}, defined as follows:

\begin{definition}
    The \textbf{P-Impact Index} of an ecosystem $E$ at time $t$ is the number of packages in $E$ at time $t$ that are transitively required by at least P\% of all the packages in $E$. 
\end{definition}

\begin{figure}[!htbp]
    \centering
    \includegraphics[width=\figsize]{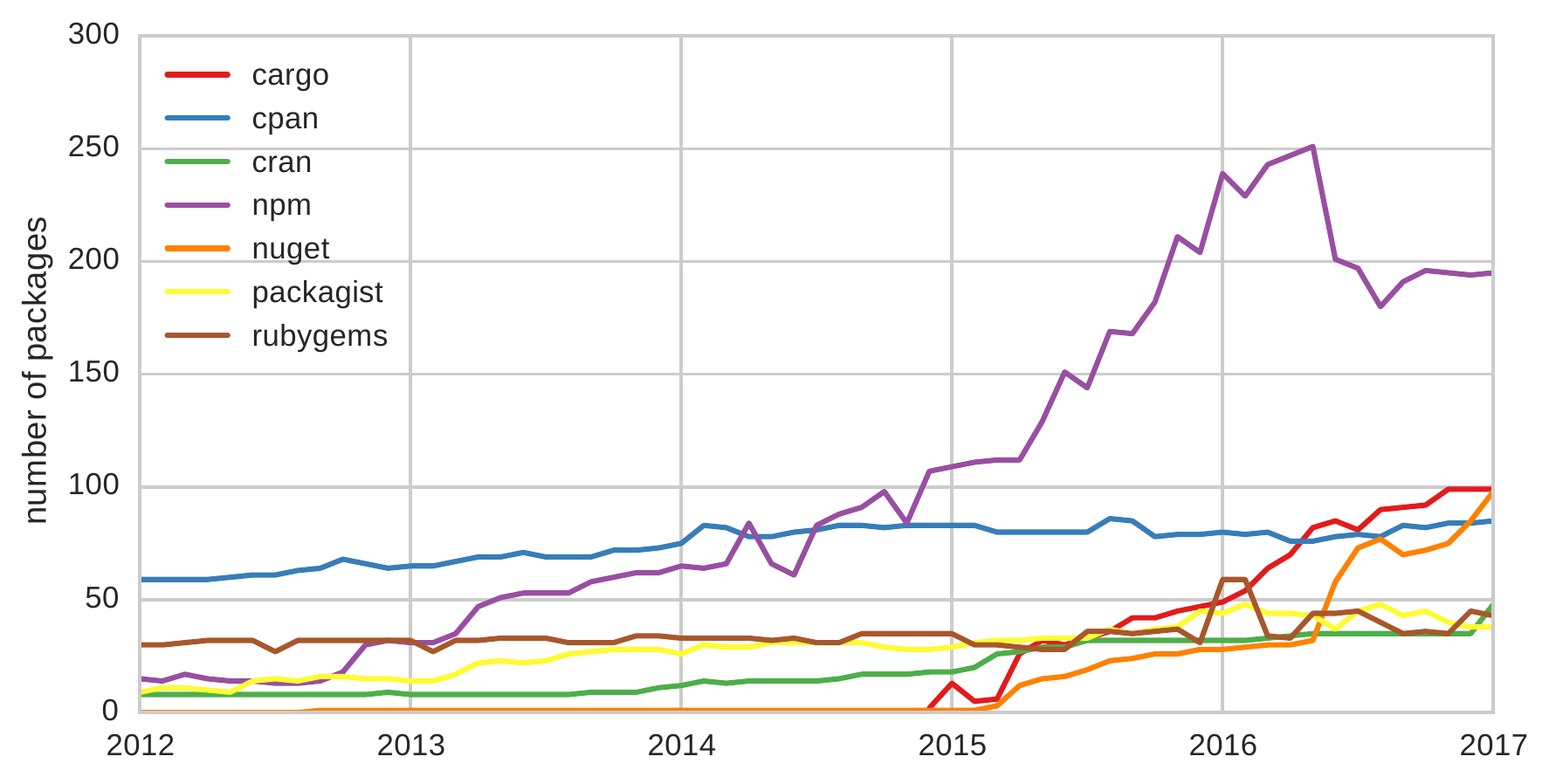}
    \caption{
      Number of packages that are transitively required by at least 5\% of all packages.
    }
    \label{fig:graph_relative_impact_5}
% min max latest
% cargo 	2.0 	99.0 	99
% cpan 	59.0 	86.0 	85
% cran 	8.0 	48.0 	48
% npm 	13.0 	251.0 	195
% nuget 	0.0 	98.0 	98
% packagist 	9.0 	48.0 	38
% rubygems 	27.0 	59.0 	43    
 \end{figure}

The P-Impact Index allows to quantify the number of packages that could have a high impact (at least P\%) in the ecosystem because of their numerous transitive dependents.
\fig{fig:graph_relative_impact_5} shows the evolution of the 5-Impact Index. The choice of 5\% was motivated by the example of the
\textsf{ActiveRecord} package in \rubygems on which relied $>5\%$ of all \rubygems packages at the time of the reported problem.
We also computed the P-Impact Index for other values of P (e.g., for 2\% corresponding to the impact of \textsf{left-pad} in \npm) and obtained similar results.

While the 5-Impact Index of \packagist and \rubygems is nearly stable over time, it continuously increases for the other ecosystems. This increase is particularly prominent for \cargo, \npm and \nuget, which also exhibit the highest values in January 2017.
For \npm, such a high Impact Index was expected, due to its large number of packages and the higher depth of their dependency tree. As for \fig{fig:graph_ratio_transitive_per_direct}, the variations observed from early 2016 onwards are most likely a reaction to the \textsf{left-pad} incident.\\
The high impact index of \cargo is somewhat surprising given its smaller size. 
While this ecosystem had only 7,421 packages in January 2017, its 5-Impact Index was already of 99, which represents more than 1.3\% of all its packages.

Based on the results of this impact analysis, we observe that 4 out of 7 ecosystems were able to restrain the fragility induced by a growing number of packages and their increasing reuse. The highest impact and growth was observed for
\npm, \nuget and \cargo, suggesting that these ecosystems should make an effort to reduce their complexity and hence their fragility.

%%%%%%%%%%%%%%%%%%%%%%%%%%%%%%

\section{Threats to Validity}
\label{sec:threats}

The metadata for all studied ecosystems was automatically gathered from \textsf{libraries.io}, with the exception of \cran where we extracted the data directly from package metadata using the \textsf{extractoR} tool\footnote{\url{https://github.com/ecos-umons/extractoR}}. Because there is no full guarantee that these tools produce correct results, we manually verified the correctness of the gathered data, and we cross-checked with other available metadata based on our previous research \cite{Decan2017SANER}, thereby reducing this threat to a minimum.

The package (release) data we used was up-to-date up to April 2017. Depending on the ecosystem's package removal policy, packages that were removed from the ecosystem before that date may have been absent from our analysis if no historical data was preserved by the ecosystem after package removal.

We constructed the dependency networks by relying on the list of dependencies explicitly provided in each package manifest. As a consequence, vendored dependencies and dynamically defined dependencies were not considered in our analyses. 
Since our collected dependencies underestimate actual reuse, we believe that this threat does not affect our results.

Some of our analyses are based on monthly snapshots of dependency networks, and we relied on the chronological order of package releases to build them. While this chronological order should match the logical order induced by the versioning scheme in most cases, this is not the case for instance for packages having multiple branches that are maintained in parallel. This is, however, unlikely to affect our findings given the scale of our analyses. 

Some analyses may be affected by local phenomena (see \fig{fig:act_volatility_index} or \fig{fig:graph_ratio_transitive_per_direct} for instance). As far as possible, we tried to pinpoint and interpret these phenomena. 
While some of these phenomena are perfectly legitimate, others are explained by a quality problem in the extracted data. For instance, the peak in the number of updates in August 2014 for \rubygems (\fig{fig:act_number_of_updates_by_month}) corresponds to the massive import of 25K package releases in \rubygems, resulting in an incorrect creation date for those package releases, which does not reflect the real date of their availability to the \ruby world. 

We do not make any claims that our results can be generalised beyond package dependency networks similar to those that we have analysed, i.e., the main package managers for specific programming languages. While the analyses that we have carried out can easily be applied to any other type of package dependency network such as \textsf{WordPress}, \textsf{Eclipse} or \textsf{Atom}, we do not expect to obtain similar results, because their packages tend to be more high-level, intended to be installed by end-users rather than be reused (through dependencies) by other packages.

%%%%%%%%%%%%%%%%%%%%%%%%%%%%%%
\section{Discussion}
\label{sec:discussion}

In Sections \ref{sec:rq-growth}~to~\ref{sec:rq-complexity} we addressed the four research questions empirically, through historical analysis of the dependency networks of seven packaging ecosystems.
This section complements this empirical analysis with additional information that may partly explain some of the observed differences.

In particular, \sect{sec:policy} discusses the effect of ecosystem-specific policies on our findings, while \sect{sec:tool-limitations}
compares the automated support for package dependency updates that has been put in place by the different ecosystems, and discusses the limitations of such support. 
Finally, \sect{sec:applications} discusses the usefulness of intergrating some of our proposed dependency network metrics into software ecosystem health analysis dashboards.

\subsection{Why Policies Matter}
\label{sec:policy}

While our empirical comparison revealed many similarities across packaging ecosystems, we also observed some important differences.
For instance, \cran features the lowest Changeability Index, one of the lowest Reusability Indices, a lower ratio of transitive to direct dependencies, and one of the lowest observed dependency depths for its top-level packages. 
This is very likely due to the fact that \cran imposes a stricter policy on its package maintainers than the other considered package managers. 

\cran follows a ``rolling release'' policy that imposes packages to be up-to-date with their dependencies~\cite{CRAN}. An automated continuous integration process based on the R CMD check tool verifies interpackage compatibilities on a daily basis. Maintainers of packages that fail the check are asked to resolve the problem before the next major \R release, and their packages get archived if they do not do so.
\cran also appears to have different evolution dynamics in many respects: despite its exponential growth, it has a lower number of monthly package updates and a corresponding higher probability of survival of package releases. \cran also witnesses a lower inequality in the distribution of package updates, resulting in a significantly lower Changeability Index.
A plausible explanation is that package maintainers are encouraged to limit the frequency of package updates to once every one or two months, in order to keep this rolling release policy manageable for package maintainers~\cite{CRAN}.

We did not find any evidence of the existence of such policies related to package updates or package dependencies for the other  ecosystems we studied. We also do not believe that those ecosystems are willing to adopt a similar process. 
Indeed, it would require package maintainers to quickly react to backward incompatible changes in package dependencies, which  represents a frequent and potentially heavy additional workload. 
This concern is shared by \cran package maintainers who consequently try to minimize or avoid dependencies on other packages, or even consider alternatives to \cran for the distribution of their packages because of this~\cite{Decan2016SANER,Bogart2015,Mens2015}.  

For the other ecosystems, the main guidelines we found do not seem to relate to package updates nor package dependencies. They rather have to do with recommendations to use semantic versioning or respecting the ``default semantics'' of the package manager. 
For instance, contrary to one's intuition, \nuget automatically selects the oldest available release that satisfies the package dependency constraints.

Another policy that may play an important role is the package removal policy. Indeed, if authors are allowed to remove their packages from the ecosystem, this increases the risk of breaking (transitive) dependents upon package removal.
Even if an ecosystem prevents packages from being removed, authors can still decide to update their packages to an empty release, leading to a potentially  similar outcome.
Some ecosystems such as \cargo or \nuget explicitly prevent packages from being removed from the ecosystem. The same is now also true for \npm, who introduced this policy as a consequence of the \textsf{left-pad} incident. However, in May 2017 \rubygems still allowed authors to easily remove their packages. In a similar vein, in May 2017  \cran still archived packages, implying that they cannot be automatically installed anymore, and thus preventing the installation of dependent packages as well. 
Hence, removal of packages can still have a high negative impact in those ecosystems.

\subsection{Limitations of Existing Support for Package Dependency Updates}
\label{sec:tool-limitations}

Di Cosmo \etal~\cite{DiCosmo2008} claims that problems related to package updates are important, and that more automated solutions to address these problems are required. This paper empirically validated these claims, by studying problems related to package updates in presence of dependent packages and by analysing how large popular packaging ecosystems currently (fail to) cope with these problems.
While we have discussed in \sect{sec:policy} how some packaging ecosystems rely on ecosystem-specific policies, let us now focus on  
technical solutions provided by each ecosystem to cope with package updates in presence of dependencies.

To avoid packages from breaking due to a dependency update, most ecosystems allow package maintainers to specify \emph{dependency constraints} on the versions of the packages they depend upon. Such constraints typically allow maintainers to explicitly select the desirable or allowed releases of a dependency, and to explicitly exclude the undesirable ones, \eg those that can contain backward incompatible changes.
While the use of dependency constraints can be beneficial to prevent backward incompatibility issues, it may as a side-effect prevent packages to benefit from updates that are released in a dependency.
This can be problematic, especially if the updates contain security or bug fixes~\cite{Cox2015}. 
A detailed empirical analysis of the use of dependency constraints was presented in \cite{Decan2017SANER} but is out of the scope of the current paper.

Combining the use of dependency constraints with \emph{semantic versioning} can enable packages to benefit from compatible updates while preventing backward incompatible ones.
Semantic versioning proposes a simple set of rules and requirements that dictate how version numbers are assigned and incremented based on the three-number version format \textsf{Major.Minor.Patch}. 
Package updates corresponding to bug fixes that do not affect the API should only increment the \textsf{Patch} version number, backward compatible updates should increment the \textsf{Minor} version number, and backward incompatible updates have to increment the \textsf{Major} version number.
Ideally, the combination of dependency constraints with semantic versioning should make it easier for package maintainers to manage dependency updates. 
Unfortunately, while it is easy to impose a semantic versioning syntax (as is the case for \cargo, \npm and \packagist), package maintainers can always decide, voluntarily or not, to break the associated versioning semantics \cite{Raemaekers2014}. 

Package maintainers can be assisted in managing their dependency updates by automated tools that monitor dependencies and notify the maintainers when a new release of a package dependency is available, or when an important update needs to be deployed.
For instance, web-based dashboards like \textsf{\url{gemnasium.com}}, \textsf{\url{requires.io}} or \textsf{\url{dependencyci.com}} provide these features as a continuous integration process, and are free to use for open source projects.
However, at the time of writing this paper, these tools monitored direct dependencies only and, therefore, did not detect update problems beyond the first level of the dependency hierarchy.

While it would be very desirable for these tools to take into account transitive dependencies as well, implementing such support is potentially very computationally expensive, especially in the presence of dependency constraints. 
Indeed, it is not unusual that several distinct releases of a same package satisfy the dependency constraints imposed by a dependent package. These releases can potentially have different lists of required packages or dependency constraints which, in turn, can potentially be satisfied by different releases, and so on, leading to an increasingly large number of potential dependency trees.
Moreover, because of transitive dependencies, a same package can be the target of different sets of dependency constraints. 
Identifying all the releases that satisfy these sets of constraints is a complex problem. 
Additionally, all considered package management systems implicitly define a conflict between any two distinct releases of a same package. This means that one cannot install two different releases of a same package, or in some cases (\cpan, \cran, \nuget and \rubygems), even two packages that transitively depend on two distinct releases of a same package.
While solutions to this problem were developed specifically for some ecosystems (\eg Debian or RPM, see~\cite{DiCosmo:2011:SCC:2025113.2025149,vouillon2013broken}), they are usually based on a SAT-solver, are not easy to implement and are potentially computationally expensive to use. 

To summarise, many techniques have been proposed and are being used in different combinations in each ecosystem to facilitate the work of package maintainers in presence of dependency updates.
Given the fact that each technique has specific drawbacks, a perfect solution does not exist.
Moreover, in addition to a good package management policy and a proper combination of the aforementioned techniques, it is essential for all package maintainers to be disciplined and act responsibly. They should limit updates to their packages, communicate with maintainers of dependent packages, limit the number of dependencies to other packages, advertise backward incompatible changes and deprecation warnings, respect the imposed policies and versioning schemes, use appropriate continuous integration and monitoring tools, and deploy bug and security fixes not only in the latest release but also in older branches.

\subsection{Towards Ecosystem-Level Health Analysis Dashboards}
\label{sec:applications}

Several dashboards for open source software development analytics are emerging. One of those is GrimoireLab\footnote{\url{http://grimoirelab.github.io}}, an open source software analytics engine commercialised by the Spanish  company Bitergia\footnote{\url{https://bitergia.com}}. Through private e-mail communication we discussed with two of Bitergia's team members about the usefulness and relevance to extend their tool suite with ecosystem-wide analyses such as those proposed in this article. They confirmed that there is indeed a need for analytics at the ecosystem level: ``\textit{[...] what we were producing was initially focused on a project and now we need to understand and provide insights about a huge amount of projects that in the end are part of an ecosystem.}'' More in particular, they agreed that there is a need for metrics that measure the health of the ecosystem, such as the ones we proposed based on the technical dependencies between software packages: ``\textit{There is a lot of interest by companies in learning about the `health' of FOSS components, and that implies learning about the components of their dependencies, and of their `siblings'. In other words, they know that the health of a single component depends on the health of the ecosystem in which it is produced and used. From the point of view of people producing software, they want to track everything around them. Just as a single example, they need to know if modules on which their software depend are healthy or not. From the point of view of users, that happens as well. For example, to understand the security problems of a product, they need to understand the security problems of all its dependencies, and in many cases, of their siblings developed by the same community.}'' 

We also discussed over e-mail with the developers of \textsf{\url{dependencyci.com}}, a continuous integration tool for monitoring package dependencies.
In particular, we asked them to share their view on the importance of package dependency networks and the potential problems caused by transitive dependencies, the focal point of our empirical analysis. They agreed that dependency-related problems tend to propagate over the dependency network:
\begin{itemize}
\item ``\textit{Whenever a measure has an upward or downward impact on its own dependencies an update in one project can cause a network-update effect that can make the whole network very noisy until it settles. Interestingly there is a direct correlation with the dependency update problem in open source that follows the same pattern.}'' 
\item ``\textit{[...] there may be a force multiplier/dampening effect up and down the tree. Relying on a project that only has one contributor, but that project is very simple and has few dependencies itself, might be acceptable. But depending upon a project that has hundreds of dependencies or security vulnerabilities and only one contributor is most likely going to cause trouble.}''
\end{itemize}
\noindent They also stressed the importance of transitive dependencies for two problems that were not specifically part of empirical analysis, namely licence compatibility and security breaches: ``\textit{transitive dependencies are incredibly useful when looking at things like licence incompatibilities. Especially when a project's more permissive licence impacts upon any of the software built upon it. Which can have direct impact up the tree. It's also useful for security notifications, some bugs will have impact on all users, regardless of where in the dependency tree the problem is.}'' 

The above discussions comfort our conviction that it is useful and relevant to integrate ecosystem-level measurements of dependency network evolution (inspired by those presented in the current article) into existing software health analysis dashboards. However, as will be discussed in \sect{sec:social}, the technical aspects of package dependencies and updates should be complemented with social aspects of developer interaction in order to come to a holistic socio-technical health analysis.

\section{Future Work}
\label{sec:futurework}

Based on the empirical analysis that we carried out and its ensuing discussion, 
this section presents a number of interesting avenues of future work.
\sect{sec:laws} postulates some laws of software ecosystem evolution that could be derived from our analysis.
\sect{sec:complexnetworks} proposes to study software ecosystems and their evolution from a complex networks perspective.
Finally, \sect{sec:social} considers to extend the technical dependency analysis with a social counterpart, by also studying the ecosystem's community of contributors.

%%%%%%%%%%%%%%%%%%%%%%%%%%%%%%
\subsection{Laws of software ecosystem evolution}
\label{sec:laws}

Lehman's famous laws of software evolution reflect established empirical observations of how individual software systems tend to evolve over time \cite{Lehman&al1997}. Based on the empirical findings in this article, we hypothesise that similar laws govern the ecosystem-level evolution of package dependency networks. Arguably the most popular laws of software evolution are the ones of \emph{Continuing Growth}, \emph{Continuing Change} and \emph{Increasing Complexity}.

If we agree to measure the size of an ecosystem's dependency network in terms of number of packages or number of dependencies, then we can claim to have found initial empirical evidence for the law of \emph{Continuing Growth} at ecosystem level, as a side-effect of answering $RQ_1$. 

We also found partial empirical evidence for the law of \emph{Continuing Change} at ecosystem level, as a side-effect of our results for $RQ_2$ where we studied the frequency of package updates, and found that the number of package updates remains stable or tends to grow over time for the studied ecosystems.
Similarly, our proposed Changeability Index was increasing over time for most of the considered ecosystems.

We also found partial support for the  law of \emph{Increasing Complexity}, if we assume that the ratio of the number of dependencies over the number of packages is an indicator of a dependency network's complexity. Another possible indicator of complexity is 
the ratio of transitive over direct dependencies, which was found to grow over time for all studied ecosystems (cf. \sect{sec:rq-complexity}). The P-Impact Index also provided evidence of an increasing fragility of the considered ecosystems.

These three laws focus on structural and technical aspects of software. Lehman has postulated other laws as well, primarily concerning the organisational and social aspects of evolving software. Since these aspects have not been part of our current empirical study, we cannot provide any initial evidence for them.
It therefore remains an open topic of future work to study to which extent Lehman's laws also hold at the level of packaging ecosystems, and whether other laws may also hold at this level.

%%%%%%%%%%%%%%%%%%%%%%%%%%%%%%
\subsection{Complex networks}
\label{sec:complexnetworks}

Complex networks are networks or graphs that contain emerging structural properties that typically do not occur in simple network structures such as lattices or random graphs \cite{Barabasi2016}. The networks of many real-world systems (e.g. the brain, social networks and computer networks) have been shown to reveal complex network properties, such as scale-freeness, the small world phenomenon, and power law behaviour. Earlier work has revealed such complex network characteristics for class dependency graphs of individual open source software systems (e.g., \cite{Myers2003Software, Zheng2008}).
Inspired by \cite{Cataldo2014}, we hypothesise that package dependency networks of open source packaging ecosystems also reveal such complex network behavior.

For example, we found a very \emph{unequal distribution of connectivity} for each ecosystem, characteristic of power law or Pareto law behaviour~\cite{Goeminne2011-SQM}. First of all, the proportion of required packages (\fig{fig:graph_prop_required_dependent_packages}) was invariably low for each ecosystem (ranging between 20\% and 30\%, and even lower for \rubygems). Secondly, a very low proportion of these required packages concentrated a very high proportion of reverse dependencies (\fig{fig:graph_required_lorenz} and \fig{fig:graph_required_gini_index}). At the other side of the spectrum we found a fairly high proportion (ranging between 40\% and 60\%) of top-level packages (i.e., connected packages that are not required by other packages) in all ecosystems. Moreover, the majority of these top-level packages had dependency trees of depth three or higher. 
We also observed a rather \emph{unequal distribution of package updates} for each ecosystem, since a major proportion of package updates was concentrated in a minority of packages (\fig{fig:act_lorenz_curve_2016}).

These initial findings make us confident that it would be worthwhile to study, compare and exploit the complex network properties of ecosystem package dependency networks as part of future work.

%%%%%%%%%%%%%%%%%%%%%%%%%%%%%%
\subsection{Socio-technical Ecosystem and Community Health Analysis}
\label{sec:social}

In the current article we have only focused on technical dependencies between packages belonging to the same ecosystem.
As explained in \cite{Mens2016keynote}, it would be very useful to study the ecosystem dynamics from a socio-technical point of view, combining information from the package dependency network with information from the social network of ecosystem contributors.

Socio-technical networks have been used frequently at the level of individual software projects, for example to predict software failures \cite{Bird2009,Posnett2013-ICSE}, to predict project or contributor abandonment \cite{Constantinou2017}, to measure successful builds \cite{Kwan2011} and many more. We are not aware, however, of any attempt to study, exploit and compare the evolution of socio-technical networks across multiple software ecosystems.

An interesting way to turn such socio-technical analysis into actionable results consists in focusing on software ecosystem and software community health aspects, by analysing and predicting social or technical events that may be detrimental to the health (e.g. quality, survival, sustainability, diversity) of the package dependency network or the social network of package contributors.
Indeed, there appears to be a general drive in the open source community to measure the health of open source communities and the software ecosystems they maintain. As an illustration of this, in September 2017, the Linux Foundation officially announced the {\textsf CHAOSS} project for Community Health Analytics of Open Source Software\footnote{\url{https://chaoss.community}}.
As part of this larger initiative, our own interuniversity {\textsf SECOHealth} project\footnote{\url{https://www.secohealth.org} (October 2017 - September 2019} will focus on understanding and assisting the health dynamics of software ecosystems.

%%%%%%%%%%%%%%%%%%%%%%%%%%%%%%

\section{Conclusion}\label{sec:conclusion}

As a follow-up on previous work \cite{Decan2017SANER}, we carried out an empirical comparison of the package dependency networks of seven packaging ecosystems, each associated to a different programming language, and available online, namely \cargo, \cpan, \cran, \npm, \nuget, \packagist and \rubygems. The range of considered ecosystems varied in size and age. Some ecosystems were very large (e.g., \npm has over 3 million package releases), while others were very old (e.g. \cpan has a release history of more than 20 years). 

The presented research is the first to compare that many different ecosystems. Previous research was limited to individual ecosystems, or at best comparison of two or three ecosystems only. In addition, the presented research is the first to use the \textsf{libraries.io} dataset containing metadata of software package dependencies of several millions of open source libraries stored in dozens of different package managers.

Our research questions related to the growth, changeability, reusability and fragility of the considered package dependency networks.

\noindent We studied the growth of package dependency networks over time, in terms of their number of packages and dependencies. We observed that these dependency networks tend to grow over time, though the speed of growth may differ. We also analysed the ratio of dependencies over packages as a simple measure of the network's complexity, and observed that this complexity either remains stable or increases over time.

\noindent We studied the changeability of package dependency networks over time, based on the number of package updates. We observed that this remains stable or tends to grow over time, and that a minority of active packages is responsible for most of the package updates. 

\noindent We studied reusability in terms of the extent to which packages depend on other packages, and oberved that dependencies are abundant in all packaging ecosystems. Most packages are connected, and this proportion increases over time. We observed that dependencies are not evenly spread across packages. A small proportion of packages concentrate a large majority of all reverse dependencies. This unequal distribution tends to increase over time. 

\noindent Finally, we studied the fragility of an ecosystem caused by the presence of transitive dependencies. We observed that a majority of dependent packages have a few direct dependencies but a high number of transitive dependencies. We identified for each ecosystem an increasing number of packages whose failure can affect an important number of other packages in the ecosystem due to transitive dependencies. 

We also contributed novel metrics, inspired by the Hirsch index, to facilitate cross-ecosystem comparison of important evolution characteristics. We defined a Changeability Index to quantify the propensity of an ecosystem to change over time, and a Reusability Index that quantifies the extent of reuse in the ecosystem. We introduced an Impact Index that quantifies the fragility of an ecosystem in terms of the number of packages having a high potential impact on the ecosystem.

We observed some important differences across ecosystems, and discussed whether and how these differences may depend on ecosystem-specific factors (such as their age, size, policies, \ldots). 
We also discussed ecosystem-specific techniques for managing package dependencies and package updates and concluded that no perfect solutions exist. We advocated the need for dependency management tools to explicitly take into account transitive dependencies, due to their prevalence and potentially high impact. 
We also advocated the need to integrate socio-technical dependency network metrics as part of software ecosystem health analysis dashboards, in order to support ecosystem managers in reducing the fragility of their ecosystems. 

\begin{acknowledgements}
This research was carried out in the context of FRQ-FNRS collaborative research project R.60.04.18.F ``SECOHealth'', ARC research project AUWB-12/17-UMONS-3 ``Ecological Studies of Open Source Software Ecosystems'', and FNRS Research Credit J.0023.16 ``Analysis of Software Project Survival''. 
We express our gratitude to Andrew Nesbitt and Ben Nickolls, both from \textsf{libaries.io} and \textsf{dependencyci.com}, for making the package manager dependency data available, and for the very useful email discussions. We thank Jesus Gonzalez-Barahona and Daniel Izquierdo from Bitergia for their relevant feedback.
We thank Eleni Constantinou, Alexander Serebrenik and Damian Tamburri for proofreading this work. 
\end{acknowledgements}

%\bibliography{biblio} 

\providecommand{\noopsort}[1]{}

\end{document}